\documentclass[12pt]{article}
\usepackage{graphicx}
\usepackage{axodraw}

\newcommand{\beq}{\begin{eqnarray}}% can be used as {equation} or
\newcommand{\eeq}{\end{eqnarray}}
\newcommand{\drawsquare}[2]{\hbox{%
\rule{#2pt}{#1pt}\hskip-#2pt%  left vertical
\rule{#1pt}{#2pt}\hskip-#1pt%  lower horizontal
\rule[#1pt]{#1pt}{#2pt}}\rule[#1pt]{#2pt}{#2pt}\hskip-#2pt%  upper horizontal
\rule{#2pt}{#1pt}}% right vertical

\newcommand{\Yfund}{\raisebox{-.5pt}{\drawsquare{6.5}{0.4}}}%  fund
\newcommand{\Yasymm}{\raisebox{-3.5pt}{\drawsquare{6.5}{0.4}}\hskip-6.9pt%
        \raisebox{3pt}{\drawsquare{6.5}{0.4}}}%  antisymmetric second rank

\newcommand{ \sla }[1]{\setbox0=\hbox{$#1$}         % set a box for #1
   \dimen0=\wd0                                     % and get its size
   \setbox1=\hbox{/} \dimen1=\wd1                   % get size of /
   \ifdim\dimen0>\dimen1                            % #1 is bigger
      \rlap{\hbox to \dimen0{\hfil/\hfil}}          % so center / in box
      #1                                            % and print #1
   \else                                            % / is bigger
      \rlap{\hbox to \dimen1{\hfil$#1$\hfil}}       % so center #1
      /                                             % and print /
   \fi}                                             %

\newcommand{\sfrac}[2]{{\textstyle\frac{#1}{#2}}}
\def\tv#1{\vrule height #1pt depth 5pt width 0pt}

\def\gtrsim{\mathrel{\mathpalette\vereq>}}
\makeatletter
\def\vereq#1#2{\lower3pt\vbox{\baselineskip1.5pt \lineskip1.5pt
\ialign{$\m@th#1\hfill##\hfil$\crcr#2\crcr\sim\crcr}}}
\makeatother

\begin{document}

\begin{titlepage}
%%%%%%%%%%%%%%%%%%%%%%%%%%%%%%%%%%%%%%%%%%%%%%%%%%%%%%%%%%%%%%%%%%%%%%%%%%%%%%%
\noindent 
\begin{flushright}
%LBNL--?????\\
UCB--PTH--01/18\\
MAD-PH-01-1227\\
{\tt hep-ph/0106044}\\
\end{flushright}

\vskip .3in
\begin{center}
{\LARGE{\bf 4D Constructions of Supersymmetric}} \\
\vskip.2cm
{\LARGE{\bf Extra Dimensions and Gaugino Mediation}} \\
%\vskip.2cm
%{\huge{\bf Gaugino Mediation}}

%\vskip .3in
%\begin{center}
%{\huge{\bf Lattice Constructions for}} \\
%\vskip.2cm
%{\huge{\bf Supersymmetric Extra Dimensions}} \\
%\vskip.2cm
%{\huge{\bf and Gaugino Mediation in 4D}}

%}}\\
%\vskip.2cm
%{\huge{\bf and 4D Gaugino Mediation from}}\\
%\vskip.2cm
%{\huge{\bf Supersymmetric Lattices}}
%\vskip.2cm
\end{center}
\vskip.2cm
\begin{center}
{\sc Csaba Cs\'aki}$^{a,}$\footnote{J. Robert Oppenheimer Fellow.},
{\sc Joshua Erlich}$^{a}$,\\
\vskip.1cm
{\sc Christophe Grojean}$^{b,c}$
and
{\sc Graham D. Kribs}$^{d}$
\end{center}

\vskip 10pt

\begin{center}
$^a${\em Theoretical Division T-8, Los Alamos National Laboratory, Los Alamos,
NM 87545}\\

\vskip 0.1in

$^b${\em Department of Physics,
University of California, Berkeley, CA 94720}\\

\vskip 0.1in

$^c${\em Theoretical Physics Group, Lawrence Berkeley National Laboratory
     Berkeley, CA 94720}\\

\vskip 0.1in

$^d${\em Department of Physics, University of Wisconsin, Madison, WI 53706}

\vskip 0.1in
{\tt  csaki@lanl.gov, erlich@lanl.gov, 
cmgrojean@lbl.gov, kribs@pheno.physics.wisc.edu}

\end{center}

\vskip .25in
\begin{abstract} 
We present 4D gauge theories which at low energies coincide with
higher dimensional supersymmetric (SUSY) gauge theories on a 
transverse lattice. We show that in the simplest case of pure 
5D SUSY Yang-Mills there is an enhancement of SUSY in the 
continuum limit without fine-tuning. This result no longer holds
in the presence of matter fields, in which case fine-tuning is necessary to
ensure higher dimensional Lorentz invariance and supersymmetry. 
We use this construction to generate 4D models which mimic
gaugino mediation of SUSY breaking.  The way supersymmetry breaking 
is mediated in these models to the MSSM is by assuming that the 
physical gauginos are a mixture of a number of gauge eigenstate gauginos: 
one of these couples to the SUSY breaking sector, while another 
couples to the MSSM matter fields.  The lattice can be as coarse as
just two gauge groups while still obtaining the characteristic 
gaugino-mediated soft breaking terms.

\end{abstract}

\end{titlepage}
%%%%%%%%%%%%%%%%%%%%%%%%%%%%%%%%%%%%%%%%%%%%%%%%%%%%%%%%%%%%%%%%%%%%%%%%%%%%%%%

%%%%%%%%%%%%%%%%%%%%%%%%%%%%%%%%%%%%%%%%%%%%%%%%%%%%%%
\section{Introduction}
\setcounter{equation}{0}
\setcounter{footnote}{0}
%%%%%%%%%%%%%%%%%%%%%%%%%%%%%%%%%%%%%%%%%%%%%%%%%%%%%%
%%%%%%%%%%%%%%%%%%%%%%%%%%%%%%%%%%%%%%%%%%%%%%%%%%%%%%
Models with extra dimensions provide several interesting mechanisms
for supersymmetry (SUSY) breaking. These mechanisms seem to make essential 
use of the presence of extra dimensions, which are not
obviously realizable in a simple four dimensional setup. Recently,
Arkani-Hamed, Cohen and Georgi~\cite{ACG} and also
Hill, Pokorski and Wang~\cite{Fermi1}
argued that it might be possible to
translate many higher dimensional effects into a purely 4D construction
by using a set of 4D theories which in the IR reproduce a
the dynamics of the extra dimensional
theory.\footnote{A similar proposal can be found in \cite{LS},
where the AdS/CFT correspondence is used to construct a purely
4D model of anomaly mediation, and in \cite{ACG2} where a higher 
dimensional stabilization of the gauge hierarchy is ``deconstructed.''
See also \cite{Fermi3}.}
These theories are also useful tools to regulate the higher dimensional
theories, and even give a UV completion of them \cite{ACG,Fermi1,Fermi2}
(see also \cite{otherlattice}).

The aim of this paper is to give a fully 4D implementation of a 
higher dimensional mechanism for supersymmetry breaking
(gaugino mediation), using 
a 4D ${\cal N}=1$ SUSY model which at low energies is equivalent to a 
latticized version of these higher dimensional
models. In order to do so we 
first show how to construct the higher dimensional supersymmetric 
theories from a 4D ``moose'' (lattice) approach. 
Because the minimal spinor representation of the 5D Lorentz group is twice as
large as that of the 4D Lorentz group,
one might think that a fine-tuning in the fermion sector is needed in order to 
construct a Lorentz invariant higher dimensional theory which includes 
fermions. 
In addition, 5D SUSY would then require at least 8 supercharges, which 
corresponds to ${\cal N}=2$ supersymmetry in 4D. 
We will demonstrate that 4D ${\cal N}=1$ supersymmetry plus gauge invariance 
(with properly chosen matter content) is enough to ensure
the existence of the additional supersymmetries in the continuum limit. 
This phenomenon of enhanced supersymmetry generation is related to the
behavior of these models at low energies in a purely 4D context, in
which ${\cal N}$=1 SUSY is enhanced to ${\cal N}$=2 on the moduli
space, without the fine-tuning of parameters.
However, in the 
presence of additional
hypermultiplets the required superpotential does have to be 
tuned in the 4D theories. The analogous effect that we obtain here is 
that maintaining 5D Lorentz invariance will require the tuning of a 
superpotential coupling in the 4D lattice models.  We present the explicit 
construction of these models which will give in the continuum 
limit the 5D ${\cal N}=1$ theory, show how to achieve the required gauge 
symmetry breaking dynamically, and how to add flavors. We carefully check
that the mass spectrum for gauge fields, scalars and fermions indeed
matches the tower of KK modes for a 5D ${\cal N}=1$ supersymmetric gauge 
theory, and that 5D Lorentz invariance and supersymmetry is indeed 
recovered in the continuum limit. This is not a trivial fact, because
in the usual Wilson lattice action the fermions are included as adjoints 
living at  
the sites of the lattice, while for our construction they are in
bifundamentals at the links.

In order to translate gaugino mediation of supersymmetry breaking into a 
4D language we show how the corresponding $S^1/Z_2$ orbifolds are constructed.
Armed with this knowledge, we present a simple 4D version of gaugino mediation,
where the lattice can be as coarse as two gauge groups and still give
the characteristic gaugino-mediated spectrum. 
One of these gauge groups 
contains the standard model matter fields, while another
couples to the supersymmetry breaking sector. The physical gaugino 
is a linear combination
of the gauginos for the various group factors
and thus obtains a mass directly from the 
supersymmetry breaking sector, while the scalar mass terms for the MSSM 
matter fields will be suppressed by an additional loop factor, just 
like in ordinary gaugino mediation. An important difference between the
4D and 5D approach is that in the 4D approach 
Planck suppressed contact terms must be subleading,
because (contrary to the 5D case) they have no further exponential 
suppression. The reason for this is that the notion of locality
from gravity's point of view is lost, if (as we will imagine) 4D gravity is 
minimally included into the theory. Thus from this point of view
the spirit of these models more closely resembles that of gauge mediation,
where the Planck suppressed operators should also be negligible.
However, the resulting mass spectrum agrees with that of gaugino mediation,
and differs from the generic gauge-mediated spectrum.

The paper is organized as follows: in Section 2 we give the construction
of the supersymmetric lattice models and check that the perturbative mass 
spectrum agrees with that of the ${\cal N}=1$ 5D theory. In Section 3
we show how to include flavors into the construction, and 
present the orbifold models. Using these results we present the 
4D models for gaugino mediation in Section 4, and conclude in Section 5.

%%%%%%%%%%%%%%%%%%%%%%%%%%%%%%%%%%%%%%%%%%%%%%%%%%%%%%
\section{The construction of supersymmetric extra dimensions}
\setcounter{equation}{0}
\setcounter{footnote}{0}
%%%%%%%%%%%%%%%%%%%%%%%%%%%%%%%%%%%%%%%%%%%%%%%%%%%%%%
%%%%%%%%%%%%%%%%%%%%%%%%%%%%%%%%%%%%%%%%%%%%%%%%%%%%%%
In Refs. \cite{ACG,Fermi1,Fermi2}, 
it has been argued that the low energy behavior of a purely 4D theory can be
effectively described by the low-lying KK modes of a 5D theory
compactified on a circle or an $S^1/Z_2$ orbifold.
Here we will first present the supersymmetric
versions of these theories, so that later we can use these to construct the
4D analogs of a higher dimensional mechanism for mediating supersymmetry
breaking.

The theory we will consider is an asymptotically free four dimensional
${\cal N}$=1 supersymmetric $SU(M)^N$ gauge
theory with chiral multiplets $Q_i$ in bifundamental representations as
follows\footnote{We will use the following conventions for our indices:
$i,j,k=1,\ldots, N$ denote the gauge group (``lattice index'') 
and until  section 3.2, we will
impose a cyclic boundary condition, {\it i.e.}, $i,j,k$ will be defined
mod $N$;
$\alpha, \beta =1,\ldots, M$ are gauge indices in the fundamental or antifundamental
representation of $SU(M)$;
and $a,b=1,\ldots, M^2-1$ are gauge indices in the adjoint representation of
$SU(M)$.}:
\begin{equation}
\label{fieldcontent}
 \begin{array}{c|ccccc}
      & SU(M)_1 & SU(M)_2 & SU(M)_3 & \cdots & SU(M)_N \\ \hline
  Q_1 & \Yfund  & \overline{\Yfund}  & 1       & \cdots & 1 \\
  Q_2 & 1       & \Yfund  & \overline{\Yfund}  & \cdots & 1 \\
  \vdots & \vdots & \vdots & \vdots & \ddots & \vdots \\
  Q_N & \overline{\Yfund} & 1 & 1 & \cdots & \Yfund 
 \end{array} \nonumber \end{equation}
The low-energy behavior of this theory
was analyzed in Ref.~\cite{CEFS}. Here we briefly
summarize the relevant results from that analysis. The flat directions 
(moduli space)
of the theory are described by the independent gauge invariant 
operators~\cite{BDFS},
which are given by $B_i={\rm det}\,Q_i,\, i=1,\dots,N$;
and $T_i={\rm tr}\,(Q_1\cdots Q_N)^i,\,i=1,\dots,M-1$. An expectation value
of the operator $B_i$ will break $SU(M)_{i}\times SU(M)_{i+1}$ to
an $SU(M)$ subgroup, leaving a theory with the same structure as the original
theory, but with one fewer $SU(M)$ factor in the gauge group. 
Once all the 
$B_i$'s have expectation values the gauge group is broken to a single
$SU(M)$. During this sequential breaking all the fields
from the first $N-1$ $Q_i$'s will become massive due to the supersymmetric
Higgs mechanism (some scalars will be eaten by the heavy gauge bosons), except 
the
fields corresponding to the trace of $Q_i$, which are then described by the
composite moduli field $B_i$. However, giving an expectation value to the 
last operator $B_N$ does not break the gauge group any further, and so 
one expects that the field $Q_N$ remains massless, and forms an adjoint 
and scalar of 
the unbroken $SU(M)$ gauge group. The invariants corresponding to the
remaining adjoint are given by the operators $T_i$ above. 
Without further modification of the model, at a generic point in
moduli space the theory will have $M-1$
unbroken $U(1)$ gauge groups and no charged fields under those $U(1)$'s.
The behavior of the gauge couplings can be described by a Seiberg--Witten
curve which has been exactly determined by considering various limits 
of the theory \cite{CEFS}.  This theory is itself an orbifold
of an ${\cal N}$=2 theory, and the dynamics of these two theories 
are closely related via the orbifold correspondence \cite{josh-asad}.

We will demonstrate that
the field theory described 
above is equivalent to a latticized version of a 5D ${\cal N}$=1 
supersymmetric gauge theory.  
This 5D ${\cal N}=1$ supersymmetric gauge theory has twice 
the number of supercharges as the ${\cal N}=1$ theory in 4D.
This is an interesting phenomenon in its
own right, as supersymmetries are dynamically generated at low energy.
Although we do not study the case here, a similar phenomenon is expected to 
occur in one dimension lower, {\em i.e.}, generation of a 4D SUSY gauge theory
from a 3D theory with fewer supersymmetry charges.

%%%%%%%%%%%%
\subsection{Dynamical generation of the symmetry breaking}

The massless matter content of the $SU(M)$ gauge theory corresponding to the
theory (\ref{fieldcontent}) at low energies
is that of an ${\cal N}=2$
4D theory, namely in addition to the massless ${\cal N}$=1
vector multiplet there is also 
a chiral multiplet in the adjoint representation. 
But in addition  the singlets $B_i$ remain massless. In order to
remove these massless fields (and at the same time provide the necessary
diagonal vacuum expectation values (vev's) of the $Q_i$'s) 
Arkani-Hamed, Cohen and Georgi
proposed the addition of a matching set of gauge singlet chiral
superfields $S_i$ and the superpotential
\begin{equation}
\label{suppot}
W_{\rm dyn.}= \frac{1}{\mu^{M-2}}\sum_i S_i (B_i- v^M),
\end{equation}
where $\mu$ is a mass scale. The question that we want to answer first is
whether this superpotential can be achieved dynamically, perhaps also
within a renormalizable theory. 
In \cite{ACG} a dynamical model for the non-supersymmetric 
case has been worked out. Here we show that for the case 
of $SU(2)$ gauge groups one can achieve this as well through supersymmetric
non-perturbative dynamics within a renormalizable theory, 
which is understood from the works of Seiberg and others \cite{Seiberg}. 
For the $SU(N)$ version of this model there will still be a 
branch on the moduli space of vacua that achieves the dynamical 
breaking of the gauge symmetry to the diagonal one. However, in order to
ensure that we are on the right branch (and that the other moduli 
are massive) a non-renormalizable tree-level superpotential will have to be
added, at least for the example based on the simplest possible matter
content. This non-renormalizable superpotential should
then be generated by some other physics at higher energies,
either through non-perturbative effects or from integrating out 
heavy particles.

We begin with an $SU(M)^{2N}$ gauge theory with the periodic structure of
(\ref{fieldcontent}).
We further assume that the gauge coupling
of every even group in the chain of $SU(N)$'s is much larger than the
neighboring odd ones: $g_2=g_{2i}\gg g_{2i-1}=g_1$, {\it i.e.},
$\Lambda_2= \Lambda_{2i} \gg \Lambda_{2i-1}= \Lambda_1$.
Therefore, concerning the dynamics of any $SU(M)_{2i}$,
the weaker gauge groups can be regarded as a weakly gauged
global symmetry, leaving  an
$SU(M)_{2i}$ gauge theory with $M$ flavors in this sector of the theory.
With this particular matter content it was shown in \cite{Seiberg} that
the theory confines in the IR with
chiral symmetry breaking, with the
confined degrees of freedom given by
$\mathcal{M}_i, \mathcal{B}_i,\tilde{\mathcal{B}}_i$ ($i=1\ldots N$),
\begin{equation}
\begin{array}{c|c|ccrc}
& SU(M)_{2i}& SU(M)_{2i-1}& SU(M)_{2i+1}& U(1)_B & U(1)_R \\
\hline
\tilde{\mathcal{Q}}_i & \overline{\Yfund}  & \Yfund & 1 & 1 & 0 \\
\mathcal{Q}_i & \Yfund  & 1 & \overline{\Yfund} & -1&  0\\
\hline
\hline
\tv{15}
\mathcal{M}_i= \mathcal{Q}_i \tilde{\mathcal{Q}}_i & &\Yfund & \overline{\Yfund} & 0 & 0\\
\mathcal{B}_i= \mathcal{Q}_i^M & & 1 & 1 & M & 0 \\
\tilde{\mathcal{B}}_i=\tilde{\mathcal{Q}}^M_i & & 1 & 1 & -M & 0 \\
\end{array} \nonumber
\end{equation}
Analyzing the 't Hooft anomaly matching conditions one concludes
that some of the global symmetries of the theory have to be broken, which
is the effect of a classical constraint on the composite fields being
modified by quantum dynamics. It was shown in \cite{Seiberg} that 
the form of the quantum modified constraint is given by\footnote{For several 
other theories with a quantum modified constraint see \protect\cite{sconf}.
For a connection between the existence of constraints among
the gauge polynomials invariants derivable from a
superpotential and the 't Hooft matching conditions see \protect\cite{syzygies}.}
\begin{equation}
{\rm det} \mathcal{M}_i -  \mathcal{B}_i \tilde{\mathcal{B}}_i = \Lambda^{2M}_2.
\end{equation}
Thus there is a branch on the moduli space where the global
$SU(M)_{2i-1}\times SU(M)_{2i+1}$ is broken to a diagonal
$SU(M)$, which is when ${\rm det} \mathcal{M}_i$ has
an expectation value. Turning on the baryons $\mathcal{B}_i$
and $\tilde{\mathcal{B}}_i$ does not break the global symmetry group.
Hence, in order to lift the branch of moduli space with
${\rm det} \mathcal{M}_i=0$
(and also to get rid of unwanted massless 
singlets) one is forced to introduce two singlets $L_i$ and $\tilde{L}_i$ and
add a tree level superpotential,
\begin{equation}
	\label{eq:supB}
W_{\rm tree} =
\frac{1}{\mu^{M-2}}(L_i \mathcal{B}_i + \tilde{L}_i \tilde{\mathcal{B}}_i),
\end{equation}
where the scale $\mu$ would have to be regarded as a cutoff scale of the
theory (perhaps originating from some other strong dynamics).
The $Q_i$'s in (\ref{fieldcontent}) should then be identified with 
the composite meson field $\mathcal{M}_i$, and
the baryons need to be lifted from the
spectrum by a superpotential term, which is generically 
non-renormalizable except for the case $M=2$ (see below).
Thus for the general $SU(M)^N$ case we do not completely
succeed in generating the model from a renormalizable 
dynamics as in the non-supersymmetric case. For this 
choice of matter content  
an extra layer of perturbative or non-perturbative dynamics 
might be needed to get
the non-renormalizable superpotentials as well.
Once this superpotential (\ref{eq:supB}) is added,
the quantum modified constraint will ensure that the remaining
$\Pi_{i=1}^{N} SU(M)_{2i-1}$
gauge groups are broken down to the diagonal $SU(M)$. In fact, 
as mentioned above, after confinement the meson matrix 
$\mathcal{M}_i$ will just play the role of the
bifundamentals $Q_i$ of (\ref{fieldcontent}) while
${\rm det} \mathcal{M}_i$ gives the invariants $B_i$, and so the
full dynamically generated superpotential in the remaining
$SU(M)^N$ theory is just,
\begin{equation}
\label{eq:sup}
\frac{1}{\mu^{M-2}} \sum_i (L_i \mathcal{B}_i + \tilde{L}_i \tilde{\mathcal{B}}_i)
+
\frac{1}{\Lambda_2^{2M-2}} \sum_i S_i (B_i- \mathcal{B}_i \tilde{\mathcal{B}}_i -\Lambda^{2M}_2),
\end{equation}
where $S_i$ are non-propagating
Lagrange multiplier chiral superfields. Integrating out the
fields $\mathcal{B}_i, \tilde{\mathcal{B}}_i$ and $L_i$
we are exactly left with the superpotential of (\ref{suppot}),
except that the symmetry breaking scale $v$ is now given by
$\Lambda_2$, and because 
the fields $Q_i=\mathcal{M}_i$ themselves are composites one gets
different powers of scales since the dimensions of these fields have not
been rescaled yet.  

The case of $SU(2)$ is special in that the tree level superpotential 
(\ref{eq:sup}) is renormalizable.  Hence, in that case the theory described
above can in fact be dynamically generated.  One might worry that because the 
representations of $SU(2)$ are pseudoreal the global symmetry in the above
analysis is enlarged from $SU(2)\times SU(2)$ to $SU(4)$ and the analysis above
would have to be modified.  
It is true that the global symmetry group is enlarged, 
but the analysis remains unchanged if we identify
${\rm det} \mathcal{M} - \tilde{\mathcal{B}}\mathcal{B}$
with the Pfaffian of the combined meson field $\mathcal{M}'$.
To be more precise, we have the following confining theory:
\begin{equation}
\begin{array}{c|c|cc}
& SU(2)& SU(4) & U(1)_R\\ \hline
  \mathcal{Q} & \Yfund  & {\Yfund} & 0\\
\hline
\hline
  \mathcal{M}'=\mathcal{Q}^2 & 1       & \Yasymm & 0 \\
\end{array} \nonumber
\end{equation}
Note that the composite meson field $\mathcal{M}'_{ij}=\mathcal{Q}_i\mathcal{Q}_j$
contains both the mesons and baryons in the previous language.  As shown in
\cite{Seiberg,IntrPoul}, 
instanton corrections will force a quantum mechanical
expectation value to the composite meson field $\mathcal{M}'$:
\begin{equation}
\label{pfaffian}
{\rm Pf} \mathcal{M}'=\mathcal{M}'_{ij} \mathcal{M}'_{kl} \epsilon^{ijkl}
=\Lambda_2^4,
\end{equation}
which will break the global $SU(4)$ to its $Sp(4)$ subgroup.
In our case, below the scale $\Lambda_2$ where the first set of $SU(2)$'s
confines the global symmetry is raised to a gauge symmetry in which case 
the effect of confinement is to break the gauged $SU(2)\times SU(2)\subset
SU(4)$ to a 
single $SU(2)$.  In the process three scalars are eaten and three remain
(of the six ``mesons'' $\mathcal{M}'_{ij}$).  Of the three massless scalars that remain,
two are given a mass by the tree level superpotential (\ref{eq:supB})
as described above,
and the Pfaffian becomes massive by the quantum modified constraint.  
The tree level superpotential and the fact that only a subgroup of the
global symmetries is gauged explicitly breaks the $SU(4)$ 
``global'' symmetry down to $SU(2)\times SU(2)$.
Alternatively, the constraint can be incorporated in the theory as before
by adding a superpotential,
\begin{equation}
S({\rm Pf}\,\mathcal{M}'-\Lambda_2^4),
\end{equation}
where $S$ is the Lagrange multiplier which enforces the constraint.
Fluctuations of the Pfaffian then obtain a mass by the Lagrange multiplier.
Hence, we have demonstrated that the supersymmetric version of the 
$SU(2)^N$ theory which is dual to the higher dimensional latticized
$SU(2)$ theory can be generated dynamically, while for larger gauge groups
one has to add a non-renormalizable superpotential if one assumes
the minimal matter content as we did here.

%%%%%%%%%%%
\subsection{Matching of the perturbative mass spectra}

Now that we understand how the model in (\ref{fieldcontent}) could
arise from supersymmetric gauge dynamics, we analyze the various mass spectra 
of the model assuming that the symmetry breaking  vev's 
%in (\ref{suppot}) 
have been generated. The aim of this analysis is
to show that one indeed recovers a higher dimensional supersymmetric
theory in the limit of $N\to \infty$, and that the number of supercharges 
is appropriately doubled. In order to be able to analyze the
massive spectrum of the model (and not just the extreme infrared like in
\cite{CEFS}) we assume that the scale $v$ is larger than the dynamical
scale of the $SU(M)$ gauge group $\Lambda < v$. This would indeed follow
in the dynamically generated examples considered above, because there
$v=\Lambda_2\gg \mu \gg \Lambda_1$. In this case the gauge
groups are broken before they could become strongly interacting, and a 
conventional perturbative analysis is possible. Then the singlet field 
corresponding to $B_i$ can be identified at the lowest order in the 
fluctuations with
$v^{N-1} {\rm tr} Q_i$, and
therefore the mass term for the
fields $S_i$ and ${\rm tr} Q_i$ from (\ref{suppot})
is given by $v (\frac{v}{\mu})^{M-2} \gg v$.

%%%%%%%%%
\subsubsection{Gauge boson masses}

\begin{figure}[tb]
$
\begin{array}{cc}
\hskip-1cm (a) & \hskip1cm (b) \\
\hskip-1cm \includegraphics*[bb=140 140 455 630,angle=-90,width=8.5cm]{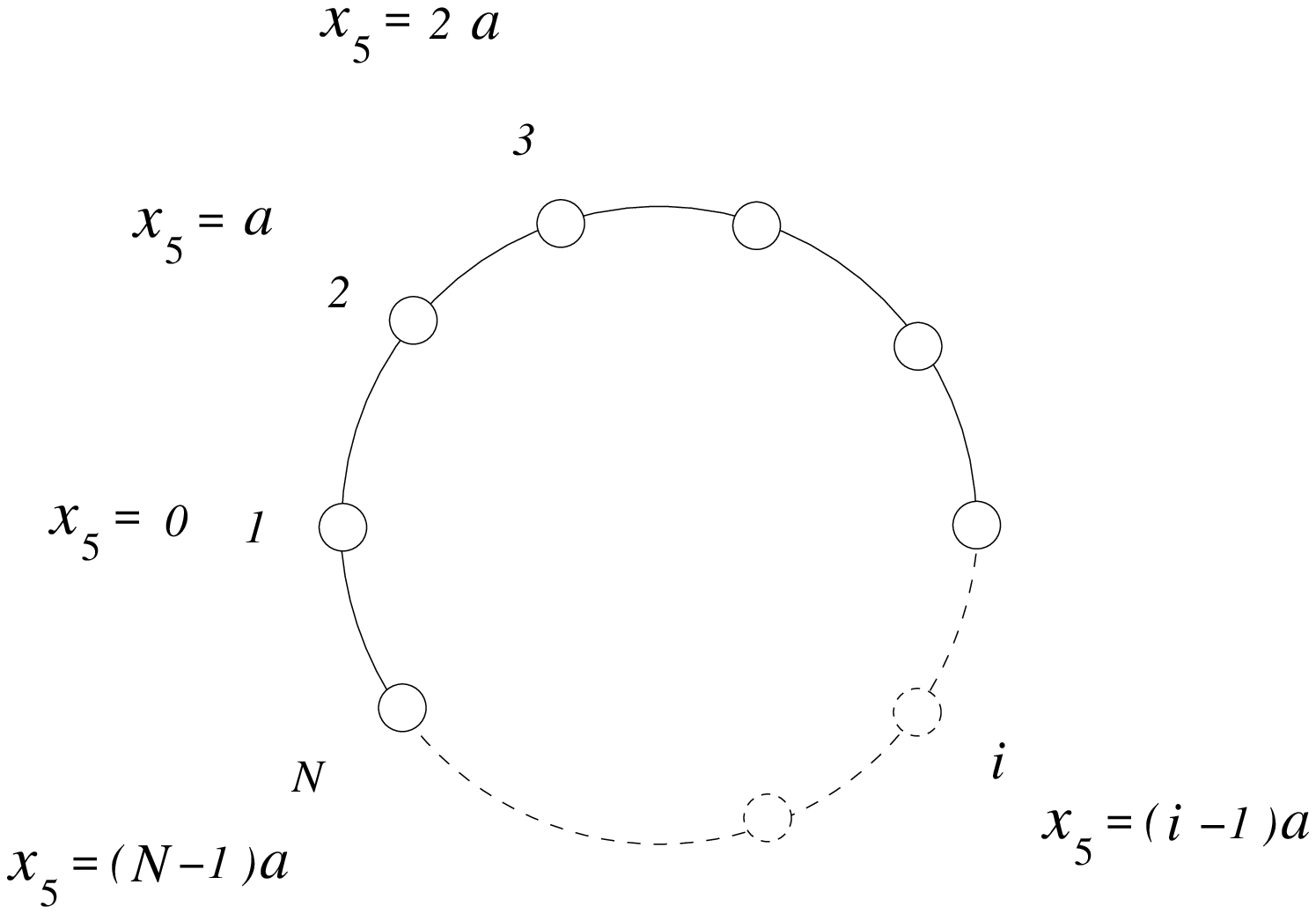}
&
\includegraphics*[bb=110 120 500 660,angle=-90,width=8.5cm]{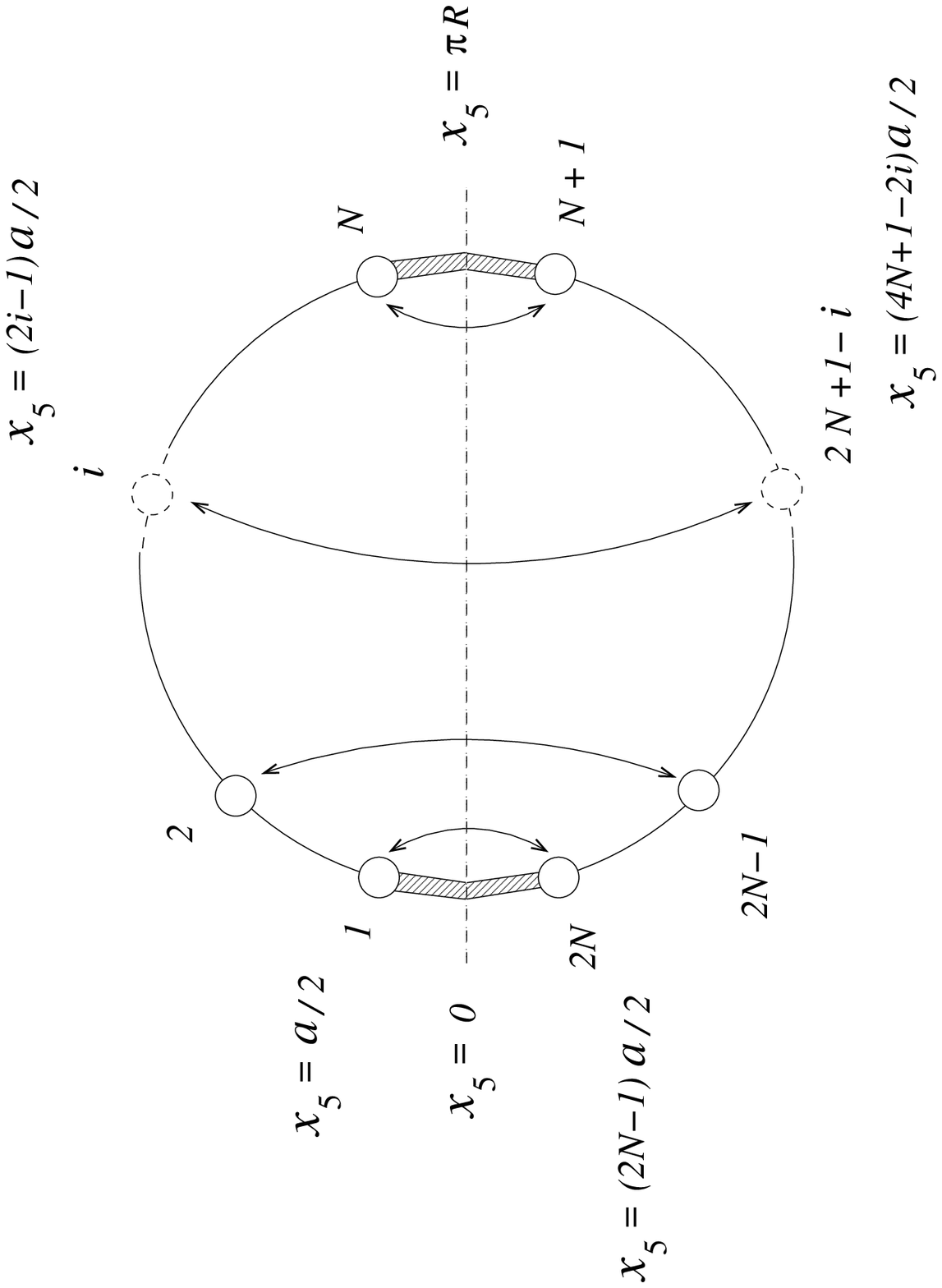}
\\
%\vspace{-.3cm}
\tilde{A}_k (x_\mu,x_5) = e^{i\, 2\pi k x_5 /Na}\, A (x_\mu)
&
\tilde{A}^+_k (x_\mu,x_5) = \cos \left( \frac{2\pi k x_5}{2 Na} \right)
\\ \\
\langle \tilde{A}_k (x_\mu) \left| A_j (x_\mu) \right. \rangle =
\sfrac{1}{\sqrt{N}}\, e^{i\, 2\pi k (j-1)  /N}
&
\langle \tilde{A}^+_k (x_\mu) \left| A_j (x_\mu) \right. \rangle =
\sqrt{\sfrac{2}{2^{\delta_{k0}}\,N}}\, \cos \left( \frac{2\pi k (j-1/2)}{2 N} \right)
\end{array}
$
\caption[]{{\small Mode decomposition for the (a) periodic and (b) orbifold
``moose" diagrams.
The mass eigenvector expansion is the discrete/latticized analogue
of the continuous Fourier expansion. The orbifold $SU(M)^N$ moose diagram is
constructed from the $SU(M)^{2N}$ periodic diagram by removing
two diametrically opposite links and identifying the sites with their reflection
about the reflecting axis.}}
\label{fig:lat}
\vskip.5cm
\end{figure}

%Next we  would like to understand the perturbative 
%mass spectrum of the theory. For this we assume that $v>\Lambda $.
The analysis of the
gauge boson mass matrix follows exactly that of the
non-supersymmetric models analyzed in \cite{ACG,Fermi1,Fermi2}, which
we repeat only for completeness. The mass matrix is obtained by expanding the 
kinetic term $\sum_i (D_\mu Q_i)^{\dagger} D^\mu Q_i$
of the scalar components of the bifundamentals $Q_i$, which
gives a contribution to the Lagrangian of the form \cite{ACG,Fermi1,Fermi2},
\begin{equation}
{\cal L}\supset
{g^2 v^2}\sum_i (A^{a\, \mu}_i-A^{a\, \mu}_{i-1})^2,
\end{equation}
where we have used the normalization ${\rm tr}\, T^a T^b =\delta^{ab}$ for the
generators of the $SU(M)$ gauge groups (this normalization will
ensure a canonically normalized kinetic term for gauginos, see later),
and $g$ is the gauge coupling
of the $SU(M)$ groups. This gives the following mass term:
\begin{equation}
\sfrac{1}{2}\, A_{i\, \mu}^a \  \mathcal{M}^2_{ij \, ab} \ A^{b\, \mu}_j
\end{equation}
where the mass matrix is a direct product of the identity in the gauge index space times
a more involved matrix in the lattice index space
\begin{equation}
	\label{gb-eq}
\mathcal{M}^2_{ij \, ab} = {2 g^2 v^2} \delta_{ab} \Omega_{ij}
\ \ \ \mbox{with} \ \ \
\Omega =
\left( \begin{array}{rrrrr}
2 & -1 & & &-1\\
-1& 2 & -1 & & \\
& \ddots & \ddots& \ddots & \\
& & -1 & 2 & -1 \\
-1 & & & -1 &2
\end{array} \right)
.
\end{equation}
The mass eigenvalues will then be given by those of the $\Omega$ matrix
with a multiplicity given by  $M^2-1$, the dimension
of the gauge index space. The diagonalization of $\Omega$ follows by
writing $\Omega$ as $2-C-C^{\dagger}$, where $C$ is the matrix of cyclic
permutations
\begin{equation}
C=\left( \begin{array}{rrrr}
 & 1 &   & \\
 &   & \ddots &\\
 &   &   & 1\\
1 &&& \\
\end{array} \right),
\end{equation}
whose eigenvectors are given by $(1,\omega_k,\omega_k^2, \ldots ,\omega_k^{N-1})$,
and eigenvalues by $\omega_k$, where
$\omega_k=e^{i\, 2\pi k/N}$, $k=0,\dots,(N-1)$, are the $N^{\rm th}$
roots of unity.
From this the mass eigenvalues are \cite{ACG,Fermi1,Fermi2}:
\begin{equation}
m_k= 2 \sqrt{2} g v \sin \frac{k \pi}{N}, \qquad 0\leq k\leq N-1.
\label{eq:mkomega}
\end{equation}
corresponding to the normalized eigenvectors:
\begin{equation}
        \label{eq:eigenbosons}
\tilde{A}^a_{k\, \mu}
=\frac{1}{\sqrt{N}}\, \sum_{j=1}^{N} \omega_k^{j-1} A^a_{j\, \mu},
\qquad 0\leq k\leq N-1
\end{equation}
This mode decomposition is just the discrete analogue
of the usual continuous Fourier expansion (see Fig.~\ref{fig:lat}(a)).
Each mass level is $M^2-1$ degenerate, forming an adjoint representation
of the unbroken diagonal $SU(M)$ gauge group
(on top of this gauge degeneracy, there is also an accidental lattice degeneracy
since $m_k=m_{N-k}$).
For small enough $k$ the spectrum approximates the Kaluza--Klein tower of
states corresponding to the compactification of the 5D theory on a circle. 
In order to find the lattice spacing of the corresponding 5D theory on a
circle with circumference $N a$,
we identify as in \cite{ACG,Fermi1,Fermi2} the low lying mass spectra
by 
\[ \frac{2\pi k}{Na}= 2 \sqrt{2} gv \frac{k\pi}{N},
\]
from which the lattice spacing is found to be 
$a=1/(\sqrt{2} gv)$.

%

%%%%%%%%%%%%%%
\subsubsection{Scalar masses}

The scalar fields in the bifundamental chiral multiplets of
(\ref{fieldcontent}) 
receive masses from the $D$-term contributions to the action.  In particular, the
Lagrangian contains a contribution
$\mathcal{L} \supset -\sfrac{1}{2}\,\sum_{a} D^a_i\,D^a_i$
for each $SU(M)_i$ factor in the gauge group, with
\begin{equation}
D^a_i = g \left(
Q_i^{*\alpha \beta} \, T^a_{\alpha\gamma} \,Q_{i}^{\gamma\beta}
-
Q_{i-1}^{*\alpha\beta} \, T^a_{\gamma\beta}\,Q_{i-1}^{\alpha\gamma}
\right)
= g\ \mbox{tr} \left(
Q_i^\dagger\, T^a \, Q_i - Q_{i-1}\, T^a \, Q_{i-1}^\dagger \right)
\, ,
\end{equation}
where $T^a$ are the generators of $SU(M)$ in the fundamental representation
(thus $-T^{a\, *}=-T^{a\, t}$ are the generators in the antifundamental) and
it is understood that we impose cyclic boundary conditions,
{\em i.e.}, $Q_{0}\equiv Q_{N}$.
When the $Q$'s develop a  vev, the fluctuations around
this  vev get a mass. Decomposing
$Q_{i}^{\alpha\beta}= v\,\delta^{\alpha\beta} + \phi_{i}^{\alpha\beta}$,
we obtain the following mass term
\begin{eqnarray}
&&
{\cal L}\supset - g^2 v^2
\sum_{i,a}
\left(
 (T^a \phi_i) (T^a \phi_i)
+  (T^a \phi_i^\dagger) (T^a \phi_i^\dagger)
+ 2 (T^a \phi_i) (T^a \phi_i^\dagger)
\right.
\nonumber\\
&&
\hskip.5cm
\left.
+(T^a \phi_i^\dagger) (T^a \phi_{i-1}^\dagger)
+(T^a \phi_i) (T^a \phi_{i-1})
+(T^a \phi_i^\dagger) (T^a \phi_{i-1})
+(T^a \phi_i) (T^a \phi_{i-1}^\dagger)
\right)
.
\label{eq:p2p1}
\end{eqnarray}
where we have defined $(T^a\phi)=\mbox{tr} (T^a\phi)$.
Using the Fierz identity for the fundamental representation of $SU(M)$
(see for example \cite{Cvitanovic}),
\begin{equation}
\sum_a T^a_{\alpha\beta}T^a_{\gamma\delta} =
\left( \delta_{\alpha\delta}\delta_{\beta\gamma}-\sfrac{1}{M} \delta_{\alpha\beta}\delta_{\gamma\delta}
\right),
\end{equation}
we obtain
\begin{equation}
	\label{Fierz}
(T^a \phi)(T^a \psi) = \mbox{tr} (\phi \psi) - \sfrac{1}{M} \mbox{tr} (\phi) \mbox{tr} (\psi)
\equiv (\phi\times\psi)
\, .
\end{equation}
Thus the mass term becomes
\begin{eqnarray}
&&
{\cal L}\supset - g^2 v^2 \sum_i \left(
(\phi_i \times \phi_i) + 2 (\phi_i \times \phi_i^\dagger)
+ (\phi_i^\dagger \times \phi_i^\dagger)
\right.
\nonumber\\
&&
\hskip2cm
\left.
- (\phi_i \times \phi_{i-1}) -(\phi_i^\dagger \times \phi_{i-1}^\dagger)
- (\phi_i^\dagger \times \phi_{i-1}) -(\phi_i \times \phi_{i-1}^\dagger)
\right)
\ ,
\end{eqnarray}
which we can rewrite as,
\begin{equation}
{\cal L}\supset -\sfrac{1}{2}\,
(\phi_{i}^{\alpha\beta} \phi_{i}^{*\alpha\beta}) \
\mathcal{M}^2_{ij\, \alpha\alpha' \, \beta\beta'} \
\left( \begin{array}{r} \phi_j^{*\alpha'\beta'} \\ \phi_j^{\alpha'\beta'} \end{array} \right)
\end{equation}
and the mass matrix is again a direct product of two matrices in the gauge
and lattice index spaces
\begin{equation}
\mathcal{M}^2_{ij\, \alpha\alpha' \, \beta\beta'}=
g^2 v^2\
\Omega_{ij}
\left( \begin{array}{rrrr}
A_{\alpha\alpha' \, \beta\beta'}  & B_{\alpha\alpha' \, \beta\beta'} \\
B_{\alpha\alpha' \, \beta\beta'} &  A_{\alpha\alpha' \, \beta\beta'}
\end{array} \right)
\ .
\end{equation}
The lattice matrix is the same as the one appearing in the gauge boson mass matrix
while now the gauge matrices are non diagonal and are given by
\begin{eqnarray}
	\label{Ascalar}
A_{\alpha\alpha' \, \beta\beta'} & = &
\delta_{\alpha\alpha'} \delta_{\beta\beta'}
-\sfrac{1}{M}\, \delta_{\alpha\beta} \delta_{\alpha'\beta'}\ , \\
	\label{Bscalar}
B_{\alpha\alpha' \, \beta\beta'} & = &
\delta_{\alpha\beta'} \delta_{\beta\alpha'}
-\sfrac{1}{M}\, \delta_{\alpha\beta} \delta_{\alpha'\beta'}
\ .
\end{eqnarray}
The second terms in (\ref{Ascalar}) and (\ref{Bscalar})
are due to the projection out of the trace in the Fierz transformations
(\ref{Fierz}), as a consequence of the tracelessness
of the generators of $SU(M)$.

We already know the eigenvalues of $\Omega$ from (\ref{eq:mkomega})
and it is easy to check
that the gauge matrix has only two eigenvalues: 0, with a degeneracy $M^2+1$
and 2, with a degeneracy $M^2-1$. The mass spectrum corresponds to the product
of these different eigenvalues as follows:
\begin{itemize}
\item $m=0$ with degeneracy $2M^2+(N-1)(M^2+1)$ ,
\item $m_k= 2 \sqrt{2} g v \sin \frac{k \pi}{N}$, for $1\leq k\leq N-1$,
with degeneracy $M^2-1$.
\end{itemize}
However, this counting of the massless modes has not taken into account the
Higgs mechanism or superpotential.
First, $(N-1)(M^2-1)$ modes are eaten in the super-Higgs mechanism
associated to the breaking of $N-1$ $SU(M)$ gauge groups, and these
would-be Goldstone modes give the longitudinal components
of the massive gauge bosons. Second, $2N$ scalars get
a mass through the $F$--terms associated to the superpotential
(\ref{suppot}). Indeed
$F_{S_i} = \mu^{-(M-2)} (\det Q_i - v^M) \supset  v (v/\mu)^{M-2} \,\mbox{tr} \phi_i$,
such that the trace of $\phi$ at each site 
acquires a large mass, $v (v/\mu)^{M-2}$, and decouples from the low energy 
effective action.
Thus we are left with only $2(M^2-1)$ real massless scalars.

%%%%%%%%%%%%%%
\subsubsection{Fermion masses}
\label{fermionmass-sec}

Finally, we consider the fermion fields.  
Our aim is to show that indeed the bifundamental fermions combine
with the gaugino to give a supersymmetric spectrum which matches that of the 
gauge 
bosons and scalars. The 4D K\"ahler potential contains
\begin{equation}
\sum_i \Phi_i^\dag\, e^{\sum_j V_j}\, \Phi_i
\end{equation}
where
$V_j$ is the vector superfield associated to the gauge group $SU(M)_j$
and
$\Phi_i$ is the link chiral superfield that transforms
as $(\Yfund,\overline{\Yfund})$ under $SU(M)_i\times SU(M)_{i+1}$.
When expanded in components, this gives the
gaugino--scalar--fermion interaction
\begin{equation}
\label{fermioninteract}
{\cal L}\supset
%\imath 
i\sqrt{2} g \sum_i \,\mbox{tr} \left(
Q_i^\dagger T^a (q_i \lambda^a_i)
- (\bar{q}_i^t \bar{\lambda}^a_i) T^a Q_i
- (q_i \lambda_{i+1}^a) T^a Q_i^\dagger
+ Q_i T^a (\bar{q}_i^t \bar{\lambda}_{i+1}^a)
\right)
\end{equation}
where $\lambda$ is the gaugino, $q$ is the two component Weyl fermion in
the bifundamental, while $Q$ is the scalar component in the bifundamental. 
Note that again these terms only give mass to the traceless parts of the
bifundamental fermions as a result of the tracelessness of the generators 
$T^a$, or equivalently because the gauginos transform in the adjoint
representation of the gauge group. Putting in the 
expectation values of the $Q$'s will give us the fermion mass terms,
which are then given by
\begin{equation}
{\cal L}\supset
%\imath 
i \sqrt{2} g v\sum_i \, \mbox{tr} \left(
\lambda_i (q_i-q_{i-1})-\bar{\lambda}_i (\bar{q}_i-\bar{q}_{i-1}) \right)
\ ,
\end{equation}
where we have defined $\lambda_{i\, \alpha\beta} = \lambda^a_i T^a_{\alpha\beta}$
(note that our normalization of the Casimir of $SU(M)$
in the fundamental representation,
{\it i.e.}, ${\rm tr}\, T^a T^b = \delta^{ab}$, ensures that
$\lambda_{\alpha\beta}$
are $M^2-1$ Weyl fermions with a canonically normalized kinetic term).
This leads to a complex mass matrix for the fermions that is once again
a direct product of a lattice and a gauge structure
\begin{equation}
\sfrac{1}{2} \, (\lambda_{i\, \alpha\beta} \left|\, q_{i\, \alpha\beta}) \right.
\ \mathcal{M}_{ij\, \alpha\alpha'\, \beta \beta'} \
\left( \begin{array}{r} \lambda_{j\, \alpha'\beta'} \\\hline  q_{j\, \alpha'\beta'} \end{array} \right)
+ {\rm h.c.}
\end{equation}
with
\begin{equation}
\mathcal{M}_{ij\, \alpha\alpha'\, \beta \beta'} = 
%\imath 
i\sqrt{2} gv \
B_{\alpha\alpha'\, \beta\beta'} \
\left( \begin{array}{r|r}  & \Theta^t \\\hline \Theta &  \end{array} \right)_{ij}
\end{equation}
where the gauge matrix $B$ has been defined in (\ref{Bscalar}) and
the lattice matrix $\Theta$ is given by
\begin{equation}
\Theta = \left( \begin{array}{rrrr}
1&-1&&\\
&\ddots&\ddots&\\
&&\ddots&-1\\
-1&&&1
\end{array} \right)
\end{equation}
It is then easy to derive the fermionic spectrum.
The square of the gauge matrix $B^\dagger B=B B^\dagger$ has one zero eigenvalue
and $M^2-1$  degenerate eigenvalues equal to 1.
On the other hand, $\Theta^\dagger\Theta= \Theta \Theta^\dagger =\Omega$, showing that the fermionic
mass levels agree with those computed before for the vectors and the scalars.
Concerning the zero modes, 
%the trace of $\lambda_i$ is projected out
as a result of the tracelessness of the generators of $SU(M)_i$
%. While, for the same reason, 
the trace of $q_i$ does not
acquire a mass from the K\"ahler potential.  Instead
it  combines with the fermionic component of the singlet $S_i$ to form
a Dirac spinor that gets a mass $v(v/\mu)^{M-2}$ from the superpotential
(\ref{suppot}). The fermionic spectrum is thus:
\begin{itemize}
\item $2(M^2-1)$ massless Weyl spinors,
\item $(M^2-1)$ Dirac spinors with mass
$m_k= 2 \sqrt{2} g v \sin \frac{k \pi}{N},\ \ 0\leq k\leq N-1$,
\end{itemize}
showing the supersymmetric nature
of the low energy spectrum.
Indeed, the massless 5D ${\cal N}$=1 vector multiplet includes a 
gauge boson
(3 components on-shell), a Dirac fermion (4 components) and a real scalar
(1 component).  Upon Kaluza--Klein reduction, we get a 4D ${\cal N}$=2 massless
vector multiplet: a gauge boson (2 components), two Weyl fermions
(2$\times$2 components) and a complex scalar (2 components); and massive vector
multiplets: massive gauge boson (3 components), fermion 
(4 components) and real scalar (1 component).  This decomposition agrees
exactly with the spectrum we have found. Moreover each mass level
transforms in the adjoint of the unbroken diagonal $SU(M)$ gauge group.

%%%%%%%%%%%
\subsection{5D Lorentz invariance and supersymmetry}

Up to now we have shown that the mass spectrum of the theory indeed
matches that of the higher dimensional supersymmetric theory.
As we have mentioned above, the question of the existence 
of the full 5D Lorentz invariance is
tightly related to the question of whether the full 5D ${\cal N}=1$
supersymmetry is present or not.
The reason is that the global ``hopping'' symmetry of the 4D theory, which
is closely related to the enhanced SUSY at low energies,
becomes a spacetime symmetry of the 5D theory.
The Lorentz symmetry generators are part of the full SUSY algebra.  Hence,
5D supersymmetry requires an enhanced Lorentz symmetry.
This manifests itself in a doubling in the number of supersymmetry generators,
which then form an irreducible representation of the 5D Lorentz group.
If supersymmetry is indeed enhanced, then the theory must
automatically be 5D Lorentz invariant for consistency
(meaning that the speed of light
should not differ in the fifth direction). The converse is also true: if one
can show that 5D Lorentz invariance is maintained, then the existence of the
four supercharges implies that there must be another set of four supercharges
present in the theory. We will pursue this latter route. We will calculate
the kinetic term along the fifth dimension for the fermionic fields,
and show that 5D Lorentz invariance is automatically obtained;
that is, the speed of light along the fifth direction automatically
matches the speed along the non-latticized four dimensions, just as 
has been suggested in \cite{ACG}. Then by
4D supersymmetry one also obtains a similar conclusion for
the scalars, from which it follows that
the full 5D supersymmetry must be present. 

In order to show this we have to show that the action
at leading order is given by terms that are the discretized versions of
the 5D ${\cal N}=1$ SUSY Yang-Mills theory. However, that theory 
has only adjoint fermions living at lattice sites, as opposed to our
bifundamental
fermions. So it is not {\em a priori} obvious that one indeed gets the required
action. In order to show this we first identify the scalar component
of the bifundamental fields with the link variable $U_{i,i+1}$ {\it via}
$Q_i=v\,  U_{i,i+1}$, where the $U_{i,i+1}$'s are unitary matrices transforming
as bifundamentals under that $SU(M)_i\times SU(M)_{i+1}$ (as is appropriate
for link variables). Now we define fermions transforming as adjoints only
under single gauge groups (thus living at the sites of the lattice in the 5D
language)
by defining $\psi_i = q_i U^\dagger_{i\, i+1}$. With this identification of the fermion fields living
on the sites we can write the interaction part of the action in
(\ref{fermioninteract}) as
\begin{equation}
\label{fermionkinetic}
i\sqrt{2} g \sum_i
\lambda_i (q_i Q^\dagger_i - Q^\dagger_{i-1} q_{i-1} )
+ {\rm h.c.}
=
i\sqrt{2} g v \sum_i
\lambda_i ( \psi_i - U_{i-1,i}^\dagger \psi_{i-1} U_{i-1,i} ) + {\rm h.c.}
\end{equation}
In 5D the Dirac fermion is irreducible, so one expects $\lambda$ and
$\psi$ to form a Dirac fermion, and the above term to correspond to
the discretized version of the kinetic term of the 5D action along the
fifth dimension. This is indeed the case, since one can define the
5D Dirac spinor by
\[ \Psi_D= \left( \begin{array}{c} i\lambda \\ \bar{\psi} \end{array} \right),\]
and then the discretized version of $i \bar{\Psi}_D \sla{D}_5 \Psi_D$
is indeed reproduced by (\ref{fermionkinetic}), where the
lattice spacing is identified with $a=1/(\sqrt{2} gv)$,
and the relevant gamma matrices in Weyl representation are\footnote{Note 
that, in order to satisfy the 5D Clifford--Dirac algebra,
the Dirac matrix in the fifth direction picks up
a factor $i$ compared to the usual $\gamma^5$ defined in 4D.}
\[ \gamma^5= i\left( \begin{array}{cc} 1\\ & -1\end{array} \right),
\ \ \gamma^0= \left( \begin{array}{cc} &-1\\ -1&\end{array} \right),
\ \ \gamma^j= \left( \begin{array}{cc} & \sigma^j\\ -\sigma^j&\end{array} \right).
\]

This shows that 5D Lorentz invariance is automatic in these models,
and in turn that the full 5D supersymmetry must be present in the
continuum limit. Thus we have shown that the kinetic terms of the
fermions automatically have the same speed of light along the 5$^{th}$
dimension as for the other four. By 4D ${\cal N}=1$ SUSY
%this implies that
the scalar kinetic terms also have the right continuum limit,
and it has already been shown in \cite{ACG,Fermi1,Fermi2} that the
same applies to the gauge bosons. In fact, as explained before,
since the 5D theory must have at least eight supercharges, there
are only two possibilities: either one obtains the 5D ${\cal N}=1$
theory and then Lorentz invariance is automatically implied, or the
theory is not Lorentz invariant. The reason for this is that the 4D
construction already guarantees the presence of four supercharges.
Thus in a Lorentz invariant theory the other four must also be present.

%%%%%%%%%%%
\subsection{Comments on non-perturbative matching}

Until now we have concentrated on the region where $v\gg \Lambda$, when
non-perturbative effects are not important, because the gauge group is
broken before it could become strongly interacting. Another important check
would be to match the non-perturbative effects for the case
when $v \sim \Lambda$. In this case non-perturbative effects will be important,
and can be described by an auxiliary Seiberg--Witten curve. These curves
have in fact been analyzed for the 4D lattice theory in \cite{CEFS},
and for the 5D ${\cal N}=1$ theories on a circle in \cite{Nekrasov}.
The degrees of both curves match, as do the number of 
moduli appearing in the theory. This suggests that there is at least a 
chance that these two curves could become equivalent in the continuum
limit. It would be very interesting to actually find a detailed
mapping of the two curves, which is however beyond the scope of this
paper.

%%%%%%%%%%%%%%%%%%%%%%%%%%%%%%%%%%%%%%%%%%%%%%%%%%%%%%
\section{Adding flavors and orbifolding}
\setcounter{equation}{0}
\setcounter{footnote}{0}
%%%%%%%%%%%%%%%%%%%%%%%%%%%%%%%%%%%%%%%%%%%%%%%%%%%%%%

%%%%%%%%%%%
\subsection{Adding flavors}
Adding extra flavors to the theory is straightforward. However, there is
one important difference compared to the case without flavors. 
Until now one did not need to
tune any parameter of the theory to recover the higher dimensional
supersymmetric model. This is not surprising, because a 4D ${\cal N}$=1
SUSY theory with only a vector and chiral multiplet
(and no superpotential for the chiral multiplet) already has ${\cal N}$=2
supersymmetry.
However, this is no longer true in the presence of hypermultiplets. In this
case the ${\cal N}=1$ Lagrangian needs to contain a superpotential
coupling of the form $\sqrt{2} g \bar{\Phi} \Sigma \Phi$, where
$g$ has to be equal to the gauge coupling, $\bar{\Phi},\Phi$ form
the hypermultiplet, and $\Sigma$ is the chiral superfield in the adjoint.
Thus we expect that a similar tuning has to occur in this case as well.
A fundamental flavor in the
5D theory will have to be included as a flavor into every
gauge group.
Thus the matter content will be modified to,
\begin{equation}
\begin{array}{c|ccccc}
  & SU(M)_1 & SU(M)_2 & SU(M)_3 & \cdots & SU(M)_N \\
\hline
  Q_1 & \Yfund  & \overline{\Yfund}  & 1       & \cdots & 1 \\
  Q_2 & 1       & \Yfund  & \overline{\Yfund}  & \cdots & 1 \\
  \vdots & \vdots & \vdots & \vdots & \ddots & \vdots \\
  Q_N & \overline{\Yfund} & 1 & 1 & \cdots & \Yfund \\
  P_1 & \Yfund & 1 & 1 & \cdots & 1 \\
  \tilde{P}_1 & \overline{\Yfund} & 1 & 1 & \cdots & 1 \\
  P_2 & 1 & \Yfund & 1 & \cdots & 1 \\
  \tilde{P}_2 & 1 & \overline{\Yfund} & 1 & \cdots & 1 \\
   \vdots & \vdots & \vdots & \vdots & \ddots & \vdots \\
  P_N & 1 & 1 & 1 & \cdots & \Yfund\\
  \tilde{P}_N & 1 & 1 & 1 & \cdots & \overline{\Yfund} \\
\end{array} \nonumber \end{equation}
The superpotential needed for this model to be ${\cal N}$=2 supersymmetric
is given by
\begin{equation}
W_{\rm flavor}= \sqrt{2} g \sum_i
{\rm tr} (\tilde{P}_i Q_i P_{i+1}) + m_0 \sum_i P_i \tilde{P}_i.
\end{equation}
This is the most general renormalizable superpotential that one can
add to
the theory, but, as explained above the coefficient in the cubic term
has to equal the gauge coupling $g$.
This superpotential generates a mass term for the fermionic components
of the flavor fields that looks like
\begin{equation}
\mathcal{L} \supset -\sfrac{1}{2}
(p_{i\alpha}\left| \, \tilde{p}_{i\alpha} )\right. \, \delta_{\alpha\alpha'}\,
\left( \begin{array}{r|r}
 & \Xi^t \\
\hline
\Xi &
\end{array}
\right)_{ij}\,
\left( \begin{array}{r}
p_{j\alpha'}\\ \hline \tilde{p}_{j\alpha'} \end{array} \right)
+ {\rm h.c.}
\end{equation}
with
\begin{equation}
\Xi=
\left( \begin{array}{rrrr}
m_0 & \sqrt{2} g v&\\
&\ddots&\ddots&\\
&&\ddots& \sqrt{2} g v\\
\sqrt{2} g v&&&m_0\\
\end{array} \right)
\end{equation}
The matrix $\Xi$ can be easily diagonalized by noting that it is written
in the
form $m_0 +\sqrt{2} g v C$ where  $C$ is the cyclic
permutation matrix whose eigenvalues are the $N^{th}$ roots of
unity  $\omega_k=e^{2 \pi i  \frac{k}{N}}$.
Thus the fermionic mass spectrum is given by
\begin{itemize}
\item $M$ degenerate Dirac spinors with (mass)$^2$,
$m_k^2= 2g^2 v^2 + m_0^2 +2 \sqrt{2} g v m_0 \cos 2\pi \frac{k}{N}$,
for $k=0\ldots N-1$.
\end{itemize}
Each mass level transforms as a fundamental of the unbroken
$SU(M)$ gauge group.
For an even number of lattice sites, the lowest mass level is
$m_0+\sqrt{2} g v$,
and thus only if we tune
this parameter to zero will we obtain a massless flavor in the bulk.
Clearly, by supersymmetry or direct calculation the mass spectrum of the complex
scalars will match that of the fermions, and we do not repeat the calculation
here.

This bulk mass term can
also be recovered by considering the interaction terms and how they
would arise from the latticized version of a higher dimensional
Lagrangian.
For example, the Yukawa coupling and mass term from the
superpotential will
have to reproduce the kinetic and mass terms of the higher
dimensional
Lagrangian. Again writing $Q_i=v U_{i,i+1}$, these terms can be
written
as
\begin{equation}
\mathcal{L} \supset
-\sqrt{2} g v
\sum_i \tilde{p}_i (U_{i,i+1} p_{i+1}- p_i)
-(m_0+\sqrt{2} g v) \sum_i \tilde{p}_{i} p_{i}
+{\rm h.c.},
\end{equation}
which is simply the lattice discretization of the kinetic term
$\bar{P}_D \sla{D}_5 P_D$ and of a bulk mass term
$(m_0+\sqrt{2} g v) \bar{P}_D P_D$
for the 5D Dirac spinors $P_D=\left(\begin{array}{r} p_i \\ \overline{\tilde{p}}_i
\end{array} \right)$.
We can see now that from this point of view the fine-tuned value of the
coupling in the superpotential was necessary in order to recover the
correctly normalized kinetic term in the 5D theory with the
lattice spacing $a=1/(\sqrt{2} gv)$.

Other multiplets can be introduced similarly, except that the
superpotential will in general be non-renormalizable.

%%%%%%%%%%%%
\subsection{Orbifolding}

Until now we have exclusively considered a periodic lattice
with link fields connecting all of the gauge groups.
In this model the full 4D ${\cal N}=2$ supersymmetry
is unbroken, including in the zero mode sector.  One interesting modification
is to explicitly break (at least some of) the supersymmetry via
the latticized analog of an orbifold.
The construction is remarkably simple:  identify the fields related by 
``reflection'' about the $Z_2$ symmetry of the moose circle
by cutting the moose in half and removing the link
supermultiplets between adjacent sites
corresponding to the orbifold fixed points 
(as in Fig.~\ref{fig:lat}(b)).\footnote{We thank Nima Arkani-Hamed for 
discussions on this point.}

We begin with a cyclic 4D $SU(M)^{2N}$ theory as described above, and
orbifold the
moose circle (the extra dimension) by a $Z_2$ reflection symmetry.   After
removing two diametrically opposite links,
this gives an $SU(M)^N$ theory corresponding to a
5D theory on an interval, similar to the
``aliphatic'' models considered by Cheng, Hill, Pokorski and Wang~\cite{Fermi1,Fermi2}.
The opening of the moose diagram also explicitly breaks the ``hopping''
symmetry of the lattice at the endpoints of the interval.
In particular, this would cause the endpoint gauge groups to be anomalous,
so we add $M$ chiral multiplets $\tilde{P}$ and $M$ chiral
multiplets $P$
in the antifundamental and fundamental representation of $SU(M)_1$
and $SU(M)_N$ respectively.  These fields would correspond to fields
stuck to the orbifold fixed points (``the branes'') in the higher 
dimensional language, and are also reminiscent of the Hor\v{a}va--Witten
compactification of 11D supergravity on an interval, in which case $E_8$ gauge
multiplets are forced to live on the endpoints of the interval to cancel
anomalies. The difference here is that after breaking of the 
gauge symmetries these fields can get a mass term from the superpotential
\begin{equation}
\frac{1}{\tilde{\mu}^{N-2}}
\sum_{i=1}^{M} \tilde{P}_i \prod_{j=1}^{N-1} Q_j P_i.
\end{equation}
In this superpotential, gauge indices on the superfields are contracted
so as to make a gauge singlet under the full $SU(M)^N$, and $\tilde{\mu}$
is a mass scale. 
Once the scalar components of the link multiplets $Q_1$
and $Q_{N-1}$ acquire  vev's, the $P$'s get a mass given by
$v (v/\tilde{\mu})^{N-2}$.
Thus they will have no  vev's for their scalar components.
The orbifold theory is summarized in the table below:
\begin{equation}
\label{orbifold}
\begin{array}{c|ccccc}
      & SU(M)_1 & SU(M)_2 & \cdots & SU(M)_{N-1} & SU(M)_N \\
\hline
{\vrule height 15pt depth 5pt width 0pt}
 \tilde{P}_1, \tilde{P}_2,\dots, \tilde{P}_M &
\overline{\Yfund} & 1 & \cdots & 1 & 1 \\
  Q_1       & \Yfund            & \overline{\Yfund}  & 1       & \cdots & 1 \\
  Q_2       & 1       & \Yfund  & \cdots & 1 & 1 \\
  \vdots    & \vdots & \vdots & \ddots & \vdots & \vdots \\
  Q_{N-1}   & 1 & 1 & \cdots & \Yfund & \overline{\Yfund} \\
  P_1, P_2,\dots, P_M       & 1 & 1 & \cdots & 1 & \Yfund\\
\end{array} \nonumber
\end{equation}
The resulting gauge boson, fermion, and scalar mass matrices
can be easily calculated. The mass term for the gauge boson is \cite{Fermi1,Fermi2}
\begin{equation}
\mathcal{L} \supset
\sfrac{1}{2}\, A_{i\, \mu}^a \  \mathcal{M}^2_{ij \, ab} \ A^{b\, \mu}_j
\end{equation}
with
\begin{equation}
\mathcal{M}^2_{ij \, ab} =
2 g^2 v^2\,
\delta_{ab}\,
\hat{\Omega}_{ij},
\ \
\hat{\Omega}=
\left( \begin{array}{rrrrr}
   1 & -1      &        &        &  \\
  -1 &  2      & -1     &        &  \\
     & \ddots  & \ddots & \ddots &  \\
     &         & -1     &   2   & -1\\
     &         &        &   -1  &  1
\end{array} \right).
\label{eq:orbgauge}
\end{equation}
The mass spectrum is \cite{Fermi1,Fermi2}
\begin{equation}
m_k^2 = 8 g^2 v^2 \sin^2 \frac{k \pi}{2 N}, \quad 0 \le k \le N - 1 .
\label{eq:orbgbmass}
\end{equation}
The zero mode remains,
%while the KK tower masses are shifted
%relative to those from a periodic lattice.
as well as half the massive modes, corresponding to the symmetric modes about
the orbifold action.
We can see that these are in fact the symmetric modes by diagonalizing the mass
matrix (\ref{eq:orbgauge}).  The resulting eigenvectors corresponding to
the modes of the 5D gauge boson are \cite{Fermi1,Fermi2},
\begin{equation}
{\tilde A}_k
=
\sqrt{\frac{2}{2^{\delta_{k0}} N}}\sum_{j=1}^{N}\cos\frac{(2j-1)k\pi}{2N}
A_{j},\qquad k=0,\dots,N-1.
\end{equation}
Note also that the wavefunctions of the periodic theory with $2N$ sites
are given by (\ref{eq:eigenbosons}),
${\tilde A}_k^{\rm periodic}= \sum_{j=1}^{2N} \omega_k^{j-1} A_{j}$
where $\omega_k$ is $e^{i\, 2\pi k/2N}$. The modes
specified by $k$ and $N-k$ are degenerate and correspond to right and left
moving modes.  The orbifold has then picked out modes with definite 
parity under
the $Z_2$ orbifold symmetry, in this case even parity.
The mode decomposition is again the discrete analogue of
the continuous orbifold expansion (see Fig.~\ref{fig:lat}(b)).

The lattice spacing
of the corresponding 5D theory on a $S_1/Z_2$ orbifold of length
$N a$ can be obtained in the same way as before, by identifying the
low lying mass spectra
\begin{equation}
\frac{\pi k}{N a} = 2 \sqrt{2} g v \frac{k \pi}{2 N} \; .
\end{equation}
We obtain $a = 1/(\sqrt{2} g v)$, the same spacing as the
periodic lattice.

We also compute the fermion masses in order to check that the spectrum
corresponds to that of the orbifolded theory.  Following
the discussion of Sec.~\ref{fermionmass-sec},
the complex fermion mass matrix in two component Weyl notation is
\begin{equation}
{\cal L} \supset
\sfrac{1}{2}  (\lambda_{i} \, q_{\hat{\imath}} )\,
\mathcal{M}_{ij}\,
\left( \begin{array}{c} \lambda_j \\ q_{\hat{\jmath}} \end{array} \right)
+ {\rm h.c.}
\end{equation}
with $i,j=1,\ldots, N$, $\hat{\imath},\hat{\jmath}=1,\ldots, N$ and
\begin{equation}
\mathcal{M}_{ij} = i \sqrt{2} gv
\left(
\begin{array}{r|r} & \hat{\Theta}^t \\
\hline
\tv{14}\hat{\Theta} & \end{array}
\right)_{ij}
\end{equation}
we have not written the gauge structure which is identical
to the cyclic case and the lattice substructure, $\hat{\Theta}$, is given by
the $(N-1)\times N$ matrix
\begin{equation}
\hat{\Theta} = \left(
\begin{array}{rrrrr}
1 & -1 \\
& \ddots & \ddots \\
&& \ddots & -1 \\
&&&1&-1
\end{array}
\right)
\label{thetahat}
\end{equation}
Although not identical to the gauge boson (mass)$^2$ matrix,
the (mass)$^2$ matrix for the fermion is closely related
\begin{equation}
\mathcal{M}^\dagger\mathcal{M} = 2 g^2 v^2
\left( \begin{array}{r|r} \hat{\Theta}^t \hat{\Theta} & \\
\hline
\tv{14}  &   \hat{\Theta} \hat{\Theta}^t
\end{array}
\right)
\end{equation}
where $\hat{\Theta}^t \hat{\Theta}$ and $\hat{\Theta} \hat{\Theta}^t$,
respectively a $N\times N$ and $(N-1)\times (N-1)$ matrix, are equal to
\begin{equation}
\hat{\Theta}^t \hat{\Theta} = \hat{\Omega},
\quad
\hat{\Theta} \hat{\Theta}^t =
\left( \begin{array}{rrrr}
2 & -1 & & \\
-1 & \ddots & \ddots & \\
&\ddots&\ddots&-1\\
&&-1&2
\end{array} \right).
\end{equation}
The eigenvalues of $\hat{\Omega}$ are given in (\ref{eq:orbgbmass}).
It can be immediately checked that $\hat{\Theta} \hat{\Theta}^t$
does not have a zero mode by evaluating its determinant
\begin{equation}
\det \, \hat{\Theta} \hat{\Theta}^t  = N \; ,
\end{equation}
In fact, the eigenvalues can be readily determined by first diagonalizing
the mass matrix
with diagonal components missing, and then adding back
the term proportional to the identity matrix.
The result is,
\begin{equation}
m_k^2=8 g^2v^2 \sin^2\left(\frac{k\pi}{2N}\right),\ 1\leq k\leq N-1.
\label{eq:orbfermmass}
\end{equation}
Therefore the massive eigenvalues pair up to give a Dirac mass term
while a Weyl fermion remains massless.
As we will see in Sec.~\ref{sec:gm} the eigenvectors of 
$\hat{\Theta} \hat{\Theta}^t$ are odd about the $Z_2$ symmetry, which specifies
the orbifold action on the link field fermions in the 5D language.

The same arguments can be applied to the scalars from
the link multiplets.  From the $D$--terms in the Lagrangian we find that
the $2(N-1)M^2$ real scalars in $Q_i$ consist
of one set of would-be Goldstone bosons that are eaten by
the $(N-1)(M^2-1)$ massive vector fields, $2(N-1)$ singlets that are given mass
by the tree level superpotential $S_iB_i$, and a set of $(N-1)(M^2-1)$
massive scalars with masses identical to the
gauge bosons.  So, we obtain a massless vector and
massless Weyl fermion, corresponding to an unbroken
4D ${\cal N}=1$ vector supermultiplet, plus a massive tower
of states that fall precisely into ${\cal N}=2$ vector
supermultiplets. We see again that the diagrammatic picture of a linear
set of gauge groups connected by link fields physically
and intuitively becomes a latticization of the line segment
obtained from a $S_1/Z_2$ orbifold.  In this particular construction 
only 4D ${\cal N}=1$ supersymmetry is preserved in the zero mode sector.

%%%%%%%%%%%%%%%%%%%%%%%%%%%%%%%%%%%%%%%%%%%%%%%%%%%%%%
%%%%%%%%%%%%%%%%%%%%%%%%%%%%%%%%%%%%%%%%%%%%%%%%%%%%%%
\section{Gaugino mediation in 4D}
\setcounter{equation}{0}
\setcounter{footnote}{0}
\label{sec:gm}
%%%%%%%%%%%%%%%%%%%%%%%%%%%%%%%%%%%%%%%%%%%%%%%%%%%%%%
%%%%%%%%%%%%%%%%%%%%%%%%%%%%%%%%%%%%%%%%%%%%%%%%%%%%%%

One application of our construction of supersymmetric 
extra dimensions is to
explore ways to communicate supersymmetry breaking
to the supersymmetrized standard model (MSSM).  The central problem
is to generate a supersymmetry breaking spectrum with no highly 
fine-tuned mass hierarchies, while simultaneously avoiding current 
bounds from experiment.
Generally this requires that the supersymmetry breaking sector
is well separated from the MSSM.  For example,
flavor non-diagonal contributions to squark and slepton 
mass matrices are severely constrained from experimental bounds 
on flavor changing neutral current processes.  One way to avoid
these constraints is to generate soft supersymmetry breaking scalar 
masses dominantly through gauge interactions.  This happens in
ordinary four dimensional gauge mediation 
where both gaugino and scalar masses are generated through one- and 
two-loop diagrams with ``messenger'' fields \cite{gaugemediation}.  
An alternative proposal,
called ``gaugino-mediation'' \cite{KKS,CLNP,SS}, physically separates 
the supersymmetry breaking sector across an extra dimension on $S_1/Z_2$
similarly to the ``anomaly-mediated'' models of \cite{anomalymediation}.
Direct couplings between the supersymmetry breaking fields and
the chiral matter fields are exponentially suppressed by the 
small wavefunction overlap of one on the other.  In this model
(contrary to anomaly mediation)
the gauge supermultiplets of the MSSM are placed in the 5D bulk,
coupling directly with the supersymmetry breaking fields that are 
assumed to be localized at one orbifold fixed point. 
The MSSM chiral matter lives on the other orbifold fixed point.
Once supersymmetry is broken, a large supersymmetry 
breaking mass is endowed to the gauginos while a loop-suppressed 
(flavor-diagonal) contribution is generated for the scalar 
masses at the compactification scale.  Large supersymmetry breaking
scalar masses are induced 
by ordinary 4D renormalization group evolution to the weak 
scale,\footnote{One or two orders of magnitude of RG 
evolution is sufficient \cite{KKS}.}
generating a spectrum that is similar to a ``no-scale''
supergravity model. 

Here we will use the construction of the supersymmetric extra dimensions
presented in the previous sections to ``translate'' the
mechanism of gaugino mediation into a purely 4D model, that will
result in a perturbative SUSY breaking soft mass spectrum identical to that of
gaugino mediation.  However, since gravity in this construction is not 
made higher dimensional, one has to ensure that the flavor changing 
Planck suppressed contact terms are subdominant.  This can be done 
by requiring that the scale of mediation of SUSY breaking $\Lambda$ 
is smaller than the Planck scale, $\Lambda \ll M_{\rm Pl}$,  
just like in gauge mediation models.  This will imply that the 
gravitino is the lightest supersymmetric particle (LSP), 
which avoids the possibility of a cosmologically troubling 
stau LSP that can occur in continuum gaugino mediation
when the size of the extra dimension is of order or smaller
than the inverse GUT scale.  In fact, lowering the SUSY mediation
scale in our case is analogous to increasing the size of the extra 
dimension in continuum gaugino mediation.

%One may ask how coarse a lattice we can take.  We will show
%that just a few lattice sites suffice, and thus no real KK tower 
%or power-law running 
%of gauge couplings occurs in this case. 
%This two lattice site model is compact and simple with
%no strong restrictions on the form of the messenger sector. 
%We need only ensure that certain $1/\Lambda$ operators
%are suppressed, which leads to a small amount of fine-tuning.

%%%%%%%%%%%%
\subsection{Gaugino masses}

We start with the setup in (\ref{orbifold}), where the $SU(M)_i$ gauge
groups are all identified with the gauge groups of the MSSM.
This generates a tower of $N-1$ states in 
massive ${\cal N}=2$ vector supermultiplet representations,
which for small $k \ll N$ are indistinguishable from the KK tower
generated in gaugino mediation.  On the $i=0$ endpoint of the
lattice, we place the sector of fields needed to break 
supersymmetry dynamically.  Rather than specifying this in detail, 
we follow Refs.~\cite{KKS,CLNP} and simply assume that the result 
of dynamical supersymmetry breaking 
is that the auxiliary component of a gauge singlet chiral superfield 
located on the $i=0$ lattice site acquires a  vev,
$\langle S \rangle = F_S \theta^2$.  The chiral matter multiplets 
of the MSSM are placed on the $i=N-1$ lattice site.  The resulting
matter content is given by
\begin{equation}
 \begin{array}{c|cccccccc}
        & SU(5)_0 & SU(5)_1 & \cdots & SU(5)_{N-2} & SU(5)_{N-1} \\ \hline
\tv{15} \tilde{P}_1, \ldots, \tilde{P}_5
        & \overline{\Yfund} & 1 & \cdots & 1 & 1 \\
           Q_1 & \Yfund  & \overline{\Yfund} & \cdots & 1 & 1 \\
        \vdots & \vdots & \vdots & \ddots & \vdots & \vdots \\
         Q_{N-1} & 1 & 1 & \cdots & \Yfund & \overline{\Yfund} \\
P_1, \ldots, P_5 & 1 & 1 & \cdots & 1 & \Yfund \\
\overline{\bf 5}_{1,2,3} & 1 & 1 & \cdots & 1 & \overline{\Yfund} \\
{\bf 10}_{1,2,3} & 1 & 1 & \cdots & 1 & \Yasymm \\
  H_d & 1 & 1 & \cdots & 1 & \overline{\Yfund} \\
  H_u & 1 & 1 & \cdots & 1 & \Yfund \\
 \end{array} \nonumber 
\end{equation}
We have written the interactions in $SU(5)$ language for
compactness, although we could also have simply latticized 
the SM gauge group.  The action on the $i=0$ point is
assumed to have the superpotential terms
\begin{equation}
{\cal L} \; = \; \int d^2 \theta\, \frac{S}{\Lambda} W_\alpha W^\alpha 
+ {\rm h.c.} \label{S-op-eq}
\end{equation}
where $W_\alpha$ is the field strength chiral superfield
for $SU(5)_0$, and $\Lambda$ is the SUSY mediation scale.
We will assume that the scale $\Lambda$ arises from supersymmetry
breaking and is larger than the inverse lattice spacing.
We have assumed for simplicity that there is a chiral multiplet $S$ with a
SUSY breaking  vev, but of course in a more complete model 
the SUSY breaking
has to be specified. For example, one could imagine that the operator  
(\ref{S-op-eq}) is generated by a gauge mediation from the SUSY breaking sector
to the gaugino, and in this case there would be an additional loop factor
$g^2/(16\pi^2)$ appearing in the gaugino mass. For more
on this possibility see Section \ref{gm}.
Once $S$ acquires a supersymmetry breaking  vev, a gaugino mass
is generated for the gaugino fields of the $SU(5)_0$ gauge group
\begin{equation}
\mathcal{L} \supset
\frac{1}{2} \frac{F_S}{\Lambda} \lambda \lambda + {\rm h.c.} \; .
\end{equation}
Below the scale $a^{-1}=\sqrt{2} g v$, the full $(2N-1)\times (2N-1)$
gaugino mass term becomes
\begin{equation}
i \frac{g v}{\sqrt{2}}
\left( \lambda_0, \lambda_1, \cdots \lambda_{N-1} |\, q_1, \cdots q_{N-1}\right)
\left( \begin{array}{ccc|ccc} 2 \epsilon_F & 0 & \cdots &&&\\
0 & 0 & & & \mbox{\Large $\hat{\Theta}$}^t&\\
\vdots & & \ddots &&&\\ \hline
&&&0& \cdots&\\ 
&\mbox{\Large $\hat{\Theta}$}&&\vdots&\ddots&\\
&&&&&\\
\end{array} \right) \left( \begin{array}{c} \lambda_0 \\ \lambda_1\\
\vdots \\ \lambda_{N-1} \\ \hline \\ q_1 \\ \vdots \\ q_{N-1}
\end{array}\right) + 
{\rm h.c.}
\end{equation}
where $\epsilon_F \equiv -i a F_S/(2 \Lambda)$, and $\hat{\Theta}$ is
given in (\ref{thetahat}).
For $|\epsilon_F| \ll 1$, the mass matrix can be approximately
diagonalized using perturbation theory.
% which is however  
%slightly tricky since t
The perturbation for the square of the mass matrix
has the following form: 
\begin{equation}
\delta\,  (M^\dagger M) = 2 g^2 v^2 \left( \begin{array}{ccc|ccc}
2 |\epsilon_F|^2 &0& \ldots & \epsilon_F^* & 0 & \ldots \\
0 & & & 0 & \\
\vdots & &\ddots & \vdots & \\ \hline
\epsilon_F & 0 & \ldots \\
0 & & & \\
\vdots & & & & \end{array} \right)
\end{equation}
Since the first order perturbation for the zero mass eigenvalue gives
a result of order $|\epsilon_F|^2$, one is forced to look at the
second order perturbations as those will also involve terms of order
$|\epsilon_F|^2$. For this, we need the eigenvectors of the unperturbed
(mass)$^2$ matrix
\begin{equation}
	\label{eq:m1/2}
2 g^2 v^2 \left( \begin{array}{c|c}
\hat{\Theta}^t \hat{\Theta} & \\
\hline
\tv{14}
& \hat{\Theta} \hat{\Theta}^t
       \end{array} \right) \; .
\end{equation}
The $N\times N$ $\hat{\Theta}^t \hat{\Theta}$ block has the following 
eigenvectors:
\begin{equation}
	\label{eq:l+}
\tilde{\lambda}^+_k = \sqrt{\frac{2}{2^{\delta_{k0}} N}} 
                      \, \sum_{m=0}^{N-1}
                      \cos \frac{(2 m + 1) k \pi}{2 N} \ \lambda_m
\qquad k = 0, \ldots, N-1
\end{equation}
with eigenvalues
\begin{equation}
m_k^2 = 8 g^2 v^2 \sin^2 \frac{k \pi}{2 N},
\qquad k = 0, \ldots, N-1 \; .
\end{equation}
The $(N-1)\times (N-1)$ $\hat{\Theta} \hat{\Theta}^t$ block has eigenvectors
\begin{equation}
	\label{eq:l-}
\tilde{\lambda}^-_k = \sqrt{\frac{2}{N}} \, \sum_{m=1}^{N-1}
                      \sin \frac{m k \pi}{N} \ q_m
\qquad k = 1, \ldots, N-1
\end{equation}
with eigenvalues
\begin{equation}
m_k^2 = 8 g^2 v^2 \sin^2 \frac{k \pi}{2 N}
\qquad k = 1, \ldots, N-1 \; .
\end{equation}
The first order perturbation for the zero mode gives a shift in
the (mass)$^2$ of $8 |\epsilon_F|^2 g^2 v^2/N$, while the
second order perturbation gives
\begin{equation}
-\frac{4}{N^2} |\epsilon_F|^2 g^2 v^2
\sum_{j=1}^{N-1} \frac{\sin^2 \frac{j \pi}{N}}{\sin^2 \frac{j \pi}{2 N}}\ .
\end{equation}
Using the relation
\[
\sum_{j=1}^{N-1} \frac{\sin^2 \frac{j \pi}{N}}{\sin^2 \frac{j \pi}{2 N}}=
2(N-1),
\]
we obtain for the full perturbation in the mass of the zero mode,
\begin{equation}
m_0 = 2 \sqrt{2} g v \frac{|\epsilon_F|}{N}.
\end{equation} 
One can also calculate the mass splittings of the higher mass fermionic
modes; however, due to the degeneracy of the mass eigenvalues one has to
use degenerate perturbation theory. The result we obtain for the 
splittings is
\begin{equation}
m_k^2=8 g^2 v^2
\left( \sin \frac{k \pi}{2 N}
\pm 2\frac{|\epsilon_F|}{N} \cos^2 \frac{k\pi}{2N} \right)
\sin \frac{k \pi}{2 N}
\qquad k = 1, \ldots, N-1 \; .
\end{equation}
All of the gaugino masses are shifted relative to the gauge
boson masses --- supersymmetry is broken!  

The zero mode gaugino mass can be written in a somewhat more 
suggestive form
\begin{equation}
m_0 = \frac{1}{N} \frac{F_S}{\Lambda} \; . \label{gaugino-mass-eq}
\end{equation}
The gaugino mass appears to vanish in the large $N$ limit.
In fact, a similar phenomenon is also present in continuum
gaugino mediation, where the corresponding
gaugino mass was given by $m_0=F_S/M^2L$.  There $M$ represented
both the scale suppressing the higher dimensional SUSY breaking 
operator as well as the scale where the 5D theory was 
becoming strongly coupled.  These scales were taken to be
equal for simplicity \cite{KKS}.  We can relate this result
to the gaugino mass found in our construction by first identifying 
one power of $1/M$ as the scale $1/\Lambda$ suppressing the 
gaugino mass operator.
The other factor of $1/M$ should be identified with the inverse 
lattice spacing. The reason is that even though the full latticized
theory is never strongly coupled (which was one of the main motivations
for this construction), below the scale of the lattice spacing 
the unbroken diagonal subgroup 
is as strongly coupled as the continuum theory 
for the same number $N$ of massive ``KK'' modes.  This suggests the 
identification  of the 
5D strong coupling scale (the other $M$) with the inverse 
lattice spacing $a$.  Then, continuum gaugino mediation \cite{KKS,CLNP} 
predicts a gaugino mass identical to Eq.~(\ref{gaugino-mass-eq}), 
\begin{equation}
\frac{1}{M L} \frac{F_S}{M} \sim \frac{1}{a^{-1} N a} \frac{F_S}{\Lambda} \; .
\end{equation}
In 5D it is easy to see that $N$ cannot be arbitrarily large 
before the gauge couplings blow up, so no consistent ``large N'' 
limit can be taken.  In our latticized theory we see the same effect
for diagonal subgroup, except that in this construction the strong 
coupling physics is resolved at the scale of the lattice spacing --- meaning
it is really an artifact of considering just the diagonal subgroup.
Above the lattice spacing the full asymptotically free product 
gauge theory is resolved, leading to a fully perturbative theory.

%%%%%%%%%%%%%
\subsection{Scalar masses at one-loop}

\begin{figure}[!t]
\begin{picture}(500,80)(0,0)
%  scalar masses at one-loop
  \DashLine(  140, 20 )( 190, 20 ){4}
  \Line(      190, 20 )( 270, 20 )
  \DashLine(  270, 20 )( 320, 20 ){4}
  \CArc(      230, 20 )( 40, 0, 180 )
  \PhotonArc( 230, 20 )( 40, 0, 180 ){4}{10}
  \put(207,52){\circle*{9}}
  \put(253,52){\circle*{9}}
  \Text(      132, 20 )[r]{$\tilde{f}_i$}
  \Text(      328, 20 )[l]{$\tilde{f}_i$}
  \Text(      230, 10 )[c]{$f_i$}
  \Text(      178, 42 )[c]{$\tilde{\lambda}_k$}
  \Text(      230, 72 )[c]{$\tilde{\lambda}_l$}
  \Text(      282, 42 )[c]{$\tilde{\lambda}_m$}
\end{picture}
\caption{{\small One-loop contribution to the MSSM matter scalar masses.
The $\bullet$'s on the gaugino line represent the SUSY breaking
mass insertion $F/\Lambda$.}}
\label{loop-fig}
\vskip.5cm
\end{figure}
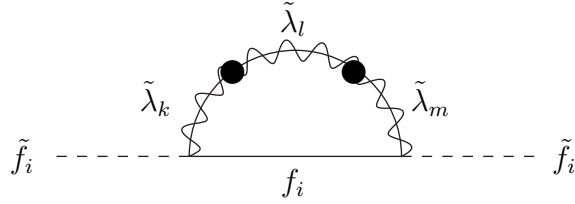

The leading contributions to the MSSM matter scalar masses 
arise through loop diagrams of gauginos interacting
with the supersymmetry breaking operators on the
$i=0$ lattice site, shown in Fig.~\ref{loop-fig}.
%%%%%%%%%%%%%%%%%%%%%%%%%%
%%%%%%%%%%%%%%%%%%%%%%%%%%%%%%%%%%%%%%%%%%%%%%%%%%%%%%%%%%%%%%%%%%%%%
The contributions are flavor diagonal, since they only involve 
some mixed combination of gauginos running in the loop.
Here we will carry out the
calculation in a way that is completely analogous to the
continuum gaugino mediation result given in Ref.~\cite{KKS}.
In particular, we will calculate the scalar mass contribution
from a one-loop diagram with a gaugino running in the loop, 
and two insertions of the non-renormalizable supersymmetry
breaking operator Eq.~(\ref{S-op-eq}).  (Hence, the gaugino
mass matrix has \emph{not} been shifted by the supersymmetry
breaking contributions since we have not yet integrated them out.)
In the interaction eigenstate basis, the MSSM matter multiplets 
interact only with the $SU(5)_{N-1}$ gaugino.  However, once the
gaugino mass matrix is diagonalized, chiral multiplets interact
with the entire tower of Majorana gaugino mass eigenstates
$\tilde{\lambda}_k$.

The gaugino wavefunctions correspond to the eigenvectors
(\ref{eq:l+}) and (\ref{eq:l-}) of the unperturbed $(2N-1)\times (2N-1)$ matrix
(\ref{eq:m1/2}).
Hence, the $2 N - 2$ massive gauginos that are paired with
equal but opposite in sign masses can be split into two sets 
of $N - 1$ gauginos with cosine and sine wavefunction expansions.
In the large $N$ limit the set of gaugino fields $\tilde{\lambda}^-_k$
with sine expansions do not directly couple to either the supersymmetry 
breaking fields or the MSSM matter fields, and so they will not
be needed in the calculations below.  

The scalar mass loop calculation involves a gaugino propagator
extending between two different lattice sites.  In the mass
eigenstate basis the full gaugino propagator is a sum over the 
$N$ gauginos.  We find it convenient to incorporate the ``endpoint''
lattice site couplings
\begin{equation}
\langle \tilde{\lambda}^+_j | \lambda_k \rangle = 
    \sqrt{\frac{2}{2^{\delta_{j0}}\,N}} \cos \frac{(2 k + 1) j \pi}{2 N}
\end{equation}
into the sum over the gaugino propagators.  The result is
\begin{equation}
{\cal P}(q; k,l) = \frac{2}{N} \sla{q} \sum_{j=0}^{N-1} 
    \frac{1}{2^{\delta_{j0}}} 
    \cos \frac{(2 k + 1) j \pi}{2 N} 
    \cos \frac{(2 l + 1) j \pi}{2 N} 
    \frac{1}{q^2 + \left( \frac{2}{a} \right)^2 \sin^2 \frac{j \pi}{2 N}}\ ,
    \label{kl-prop-eq}
\end{equation}
which represents the summed gaugino propagator with Euclidean
momentum $q$ extending between the $k^{\rm th}$ to $l^{\rm th}$
lattice sites.  Note that we have not written the mass term
since it will drop out of the scalar mass calculation below.
We only need
the propagator extending from the $k=0$ to $l=N-1$ lattice site.
With suitable rearrangements, this is
\begin{equation}
{\cal P}(q; 0,N-1) = \frac{2}{N} \sla{q} \sum_{j=0}^{N-1} 
   \frac{1}{2^{\delta_{j0}}}
   \frac{(-1)^j \cos^2 \frac{j \pi}{2 N}}{q^2 
   + \left( \frac{2}{a} \right)^2 \sin^2 \frac{j \pi}{2 N}} \; .
\end{equation}
We note that in the large $N$ limit, this reproduces
the continuum gaugino propagator found in Ref.~\cite{KKS}.
The finite sum can be done, and we find
\begin{equation}
{\cal P}(q; 0,N-1) = a^2 \sla{q} \prod_{j=0}^{N-1} \frac{1}{(a q)^2 
   + 4 \sin^2 \frac{j \pi}{2 N}} \; .
\end{equation}
Given this relatively simple expression for the gaugino propagator
extending between the endpoints of the lattice, we can
now carry out the scalar mass calculation.

The one-loop diagram for scalar masses can be written in terms of
the above summed gaugino propagator as
\begin{equation}
\tilde{m}^2 = \frac{g^2}{16 \pi^2} \left| \frac{F_S}{\Lambda} \right|^2
\int d^4 q \,\, {\rm tr} \left[ P_R \frac{1}{\sla{q}} P_L
{\cal P}(q; N-1, 0) P_R {\cal P}(q; 0,0) P_L {\cal P}(q; 0; N-1) \right] \; .
\end{equation}
The propagator for the zeroth to zeroth lattice site can be
obtained from Eq.~(\ref{kl-prop-eq}),
\begin{equation}
{\cal P}(q; 0,0) 
= \frac{\sla{q}}{N q^2}
\left[ 1 + \sum_{k=1}^{N-1} \frac{2 (a q)^2 \cos^2 \frac{k \pi}{2 N}}{(a q)^2
+ 4 \sin^2 \frac{k \pi}{2 N}} \right] 
\equiv \frac{\sla{q}}{N q^2}
\left[ 1 + \Sigma \right] \; .
\end{equation}
Notice that, unlike ${\cal P}(q; 0, N-1)$, there are no delicate 
cancellations between the zero mode and the massive tower states.
The scalar mass calculation therefore reduces to performing
the integral
\begin{eqnarray}
\tilde{m}^2 &=& \frac{g^2}{16 \pi^2 N} \left| \frac{F_S}{\Lambda} \right|^2
        a^4 \int d^4 q \prod_{n=0}^{N-1} 
        \frac{1}{\left[ (a q)^2 + 4 \sin^2 \frac{n \pi}{2 N} \right]^2}
        \left[ 1 + \Sigma \right] \\
&=& \frac{g^2}{16 \pi^2 N} \left| \frac{F_S}{\Lambda} \right|^2
\int_{2 \sin \frac{\pi}{2 N}}^{\infty} d (a q)\frac{2 \pi^2}{(a q) 
    \prod_{n=1}^{N-1} \left[ (a q)^2 + 4 \sin^2 \frac{n \pi}{2 N} \right]^2}
    \left[ 1 + \Sigma \right]
\end{eqnarray}
One approximation is to neglect the sum over the massive tower of 
gauginos for ${\cal P}(q; 0,0)$, i.e., $[1 + \Sigma] \rightarrow 1$.
This gives a reasonable estimate to within a factor of four or so,
however we will retain the full sum in our calculations below.
Notice that the integral is logarithmically IR divergent but UV finite,
and so we start the momentum integration at the scale of the
lightest massive gaugino.  The IR divergence is handled by the
usual 4D logarithmic evolution of the zero mode 
gaugino mass for energy scales $q < 2 a^{-1} \sin \frac{\pi}{2 N}$.

The momentum integral (including the sum) evaluates to $c/N^2$ with 
a numerical coefficient $c$ that asymptotes to $1$ to very good 
accuracy for large $N$ ($|c - 1| < 0.1$ for $N > 4$).
The gauge coupling $g$ 
for the $SU(5)_{N-1}$ group on the $N-1$ lattice site must also be
converted to the gauge coupling of the unbroken diagonal 
subgroup $SU(5)_{\rm diag}$ via
\begin{equation}
g_{\rm SM} = \frac{g}{\sqrt{N}} \; .
\end{equation}
If we ignore other factors of $2$ and quadratic Casimirs,
we obtain
\begin{eqnarray}
\tilde{m}^2 &=& \frac{g_{\rm SM}^2}{16 \pi^2} 
\left| \frac{1}{N} \frac{F_S}{\Lambda} \right|^2 \\
&=& g_{\rm SM}^2 \left( \frac{m_0}{4 \pi} \right)^2 \; . 
\label{one-loop-compact-eq}
\end{eqnarray}
This result agrees exactly with continuum gaugino mediation.
Hence, our 4D latticized supersymmetric theory generates 
a gaugino and scalar mass spectrum that is identical to the 
5D continuum gaugino mediation result.

%%%%%%%%%%%%%%
\subsection{Other nongravitational contributions to scalar masses}

Here we consider whether the induced scalar masses are 
really flavor diagonal.  In continuum gaugino mediation, it was 
argued that direct couplings between MSSM matter scalar
masses and supersymmetry breaking fields are forbidden
by 5D locality.  However, the wavefunctions of the fields 
localized to the orbifold fixed points are not truly
delta functions, but instead have some width extending into
the fifth direction.  The overlap of fields located on one
fixed point with the other is therefore anticipated to be 
exponentially suppressed by roughly $e^{-M L}$.  What is the analog 
in our construction?  Naively 4D effective theory suggests we should 
be able to write dangerous operators such as
\begin{eqnarray}
&&
\int \, d^4  \theta \, \frac{S^\dag S}{\Lambda^2} L_i^\dag L_j \label{S-bad-op}
\\
&&
\int \, d^4 \theta \, \frac{Q_k^\dag Q_k}{\Lambda^2} L_i^\dag L_j
\label{Q-bad-op}
\end{eqnarray}
where $L_i$ can be any matter superfield of the MSSM.
We assume that the same scale $\Lambda$ suppressing the gaugino 
mass operator, Eq.~(\ref{S-op-eq}), also enters these operators.
However, the coefficients are undetermined and could be order one
or (in a gauge mediation model) could be loop suppressed.
However, these operators would only be generated if the MSSM fields
coupled directly to the SUSY breaking sector. Our assumption is 
that it is only the $i=0$ gauge group that couples to SUSY breaking;
thus these operators which directly couple the MSSM to the SUSY breaking
sector are absent, due to this version of ``locality on the lattice.''
There are, however, ``local'' operators that 
contribute to flavor nondiagonal scalar
masses.  These have the form
\begin{equation}
\int d^4 \theta\, \frac{S^\dag S}{\Lambda^{N + 1}}
    {Q_1} {Q_2} \ldots Q_{N-1} L_i^\dag L_j \label{link-op-eq}
\end{equation}
After the link fields acquire  vevs, this gives rise to an
operator of the form
\begin{equation}
\frac{v^{N-1}}{\Lambda^{N-1}} \int d^4 \theta \, \frac{S^\dag S}{\Lambda^2}
    L_i^\dag L_j \; .
\end{equation}
Clearly once supersymmetry is broken this
gives rise to a flavor-nondiagonal scalar mass
\begin{equation}
\frac{v^{N-1}}{\Lambda^{N-1}} \left| \frac{F_S}{\Lambda} \right|^2 
\phi_{L_i}^* \phi_{L_j} \; .
\end{equation}
If we write $\epsilon_v \equiv \Lambda/v$, then this
contribution becomes
\begin{equation}
\epsilon_v^{-N-1} \left| \frac{F_S}{\Lambda} \right|^2 \phi_{L_i}^* \phi_{L_j} 
\; .
\end{equation}
For $\epsilon_v < 1$, this contribution is power suppressed
by the number of lattice sites $N$.  We have already
found that the relationship between the latticized theory
and the continuum is $N \sim M L$, and so we see
that the latticized theory has an analog of the exponential suppression
expected in continuum gaugino mediation.  For this to be the case, 
it is crucial that the scale suppressing this higher dimensional 
interaction $\Lambda$ is larger than the induced link  vev
(or inverse lattice spacing).

In addition, the link field scalars also acquire supersymmetry 
breaking masses.  This arises because we can write the operator
\begin{equation}
\int d^4 \theta \, \frac{S^\dag S}{\Lambda^2} Q_1^\dag Q_1
\end{equation}
for the first link field, which leads to the mass term,
\begin{equation}
\left|\frac{F_S}{\Lambda}\right|^2 \phi_1^* \phi_1 \; . \label{Q1-mass-eq}
\end{equation}
Since the link field already has a (large)  vev, this supersymmetry
breaking mass simply shifts the scalar mass by $|F_S/\Lambda|^2$
which is of order the gaugino mass.  The other link
fields do not have a direct coupling to the SUSY breaking
sector, and so acquire loop suppressed SUSY breaking contributions
for their scalar components.

%%%%%%%%%%%
\subsection{Planck suppressed contributions to scalar masses}

We argued that the operators in Eqs.~(\ref{S-bad-op}),(\ref{Q-bad-op})
are absent due to the assumed ``locality on the lattice.''
However, in our construction gravity is assumed to be ordinary 4D 
Einstein gravity, and thus we have
every reason to expect ordinary 4D Planck suppressed operators 
will violate ``locality on the lattice.''
In particular, the usual Planck suppressed operators
resulting from replacing the dynamical scale $\Lambda$ with $M_{\rm Pl}$
in Eq.~(\ref{S-bad-op}) are present here
\begin{equation}
\int d^4 \theta \, \frac{S^\dag S}{M_{\rm Pl}^2} 
L_i^\dag L_j \label{Pl-bad-op} \; 
\end{equation}
where again $L_i$ can be any matter superfield of the MSSM.
These give rise to flavor off-diagonal contributions to
scalar masses of order $|F_S/M_{\rm Pl}|^2$.  The limits
on the size of these contributions are strongly model dependent, 
but roughly one finds the ratio of off-diagonal to diagonal scalar
(mass)$^2$s to be $m_{ij}^2/m_{ii}^2 < 10^{-4}$ 
for at least some choices of $i\not=j$.  This means that
we must require
\begin{equation}
\left|\frac{F_S}{M_{\rm Pl}}\right|^2 < 10^{-4} |m_0|^2
\end{equation}
or that
\begin{equation}
N \Lambda < 10^{-2} M_{\rm Pl} \; .
\end{equation}
%
%This can be satisfied if $N$ is of order a few and $\Lambda$ does
%not exceed the gauge coupling unification scale $\sim 10^{-2} M_{\rm Pl}$.
This is a separate requirement that must be imposed on our
latticized theory (that does not appear in continuum 
gaugino mediation).  Interestingly, this also implies that the 
gravitino is the lightest supersymmetric particle since
\begin{equation}
m_{3/2} = \frac{F_S}{M_{\rm Pl}} < 10^{-2} m_0 \; .
\end{equation}

\subsection{Gauge-mediated contributions to scalar masses
\label{gm}}

\begin{figure}[!t]
\begin{picture}(500,90)(0,0)
%  h-center at 230
%  scalar masses at two-loops
  \DashLine(  170, 20 )( 230, 20 ){4}
  \DashLine(  230, 20 )( 290, 20 ){4}
  \Photon(    230, 20 )( 200, 60 ){4}{6}
  \Photon(    230, 20 )( 260, 60 ){4}{6}
  \DashCArc(  230, 30 )( 42.4 , 45, 135 ){1}
  \DashCArc(  230, 90 )( 42.4 , 225, 315 ){1}
  \Text(      162, 20 )[r]{$\tilde{f}_i$}
  \Text(      298, 20 )[l]{$\tilde{f}_i$}
\end{picture}
\caption{{\small Two-loop gauge-mediated contribution to the MSSM matter 
scalar masses (only one example diagram shown).  The internal
loop of dotted lines correspond to the messenger scalars.}}
\label{twoloop-fig}
\vskip.5cm
\end{figure}
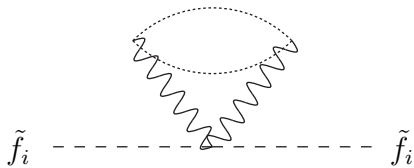

Up to now we have assumed that the only source of supersymmetry
breaking in the model is the operator in (\ref{S-op-eq}). If that is 
indeed the case, then the relevant contributions to the soft breaking
mass terms are the ones listed in the previous sections. However,
in more realistic models the operator (\ref{S-op-eq}) appears through 
gauge mediation from the messenger fields, and therefore the gaugino mass
itself has a loop suppression factor $g^2/(16\pi^2)$.
The expression for the gaugino mass is given by
\begin{equation}
m_{\rm gaugino}=\frac{g^2}{16 \pi^2 N} \frac{F_S}{\Lambda}=
\frac{g_{SM}^2}{16 \pi^2} \frac{F_S}{\Lambda},
\end{equation}
which is the ordinary 4D gauge mediation result. In this case, however, the 
MSSM scalars will also pick up a soft breaking (mass)$^2$ term from
gauge mediation, which is no longer loop suppressed compared
to the gaugino mass.\footnote{We thank Yuri Shirman for reminding us of these
operators.} An example of a two-loop diagram of this sort is given 
in Fig.~\ref{twoloop-fig}.
%%%%%%%%%%%%%%%%%%%%%%%%%%
%%%%%%%%%%%%%%%%%%%%%%%%%%%%%%%%%%%%%%%%%%%%%%%%%%%%%%%%%%%%%%%%%%%%%
These diagrams have been explicitly evaluated for 
an extra dimension on $S^1/Z_2$ with gauge fields in the bulk
by Mirabelli and Peskin \cite{MP}.  There they assumed the 
messenger sector and the MSSM matter fields were separated on
the two orbifold fixed points separated by a distance $L$ in the
fifth dimension.  They found that the gauge-mediated
contribution to the scalar mass is suppressed by an additional factor
of $1/(ML)^2$ where $M$ is the cutoff scale. 
In the latticized case we therefore expect that
the scalar mass will be suppressed by an 
additional factor of $1/(\Lambda N a)^2$ compared to the gaugino. In order
to estimate this diagram, we use the intuitive derivation given in 
\cite{MP} in which the extra loop of messengers is shrunk to a point.
It was shown in \cite{MP} that the effect of the messenger loop can be
represented as a two derivative effective operator that results in  
an extra factor of $q^2$ in the loop.
Thus we can estimate the size of the scalar masses to be of 
order\footnote{Note that the factor $1/(\Lambda a)^2$ was erroneously
omitted in (\ref{two-loop-complicated-eq}) and (\ref{two-loop-simple-eq})
in the first version of this paper.  This was corrected after the
appearance of Ref.~\cite{CKSS}.}
\begin{equation}
\tilde{m}^2 \sim \left( \frac{g^2}{16 \pi^2} \right)^2
\left| \frac{F_S}{\Lambda} \right|^2 \frac{1}{(a \Lambda)^2} 
\int_{2 \sin \frac{\pi}{2 N}}^{\infty} d (a q)\frac{2 \pi^2 (aq)}{ 
    \prod_{n=1}^{N-1} \left[ (a q)^2 + 4 \sin^2 \frac{n \pi}{2 N} \right]^2}.
    \label{two-loop-complicated-eq}
\end{equation}
We have numerically verified that the integral is well approximated
by $c/N^4$ with $c \sim 4$ at large $N$, and so the expression 
we find for the scalar masses is given by
\begin{equation}
\tilde{m}^2 \sim \frac{1}{(\Lambda N a)^2} m_{\rm gaugino}^2 \; .
    \label{two-loop-simple-eq}
\end{equation}
We can see that if $\Lambda a \sim 1$, then by decreasing $N$
the scalar spectrum interpolates between gaugino mediation for $N \gtrsim 5$
and ordinary gauge mediation for $N = 1$.
If however $1/(\Lambda a) \ll 1$, then a gaugino-mediated
spectrum is obtained even for $N=2$ \cite{CKSS}.

%%%%%%%%%%%%
\subsection{Realistic models}

We can now use these results to construct realistic models
of mediating supersymmetry breaking to the MSSM.
If a messenger sector does live on the SUSY
breaking site, we recover a 
variation of the gauge-mediated spectrum.  The important
difference between ordinary gauge mediation and our
latticized version is that there is an additional $1/(\Lambda N a)^2$ 
suppression of the scalar (mass)$^2$s relative to the gaugino (mass)$^2$.
Raising $\Lambda N a$ therefore has a similar effect on the sparticle
mass spectrum as raising the number of messenger fields in 
ordinary gauge mediation, in which the the scalar (mass)$^2$ 
are suppressed by a factor $1/n_{\rm mess}$ relative
to the gaugino (mass)$^2$.

For larger $\Lambda N a \gtrsim 5$, or for any $N$ if SUSY breaking is 
communicated exclusively by the operator Eq.~(\ref{S-op-eq}), 
the soft mass spectrum is identical to gaugino mediation. 
One interesting possibility is to consider how small $N$ can
be and yet also obtaining a viable soft mass spectrum.
If $1/(\Lambda a) \ll 1$, then one obtains the gaugino mediation
spectrum even for $N=2$.  
For remainder of this section, we wish to consider this case with 
with just two gauge groups.

\subsubsection{Two gauge groups}

The dominant contribution to the scalar masses comes from the one-loop
diagram given by (\ref{one-loop-compact-eq}).  In this case, the
the integral over momentum can be done exactly
\begin{equation}
\int_{\sqrt{2}/2}^{\infty} d (a q) \frac{2 \pi^2}{(a q) 
   \left[ (a q)^2 + 2 \right]^2} \left[ 1 + \frac{(a q)^2}{(a q)^2 + 2} \right]
= \pi^2 \frac{-3 + 8 \ln 2}{8 N^2} \sim 0.785 \; .
\end{equation}
We see that this integral evaluates to $c/N^2$ with $c \sim 3$, and 
therefore the scalar masses are slightly larger than
what would be expected for a large number of lattice sites.
They are, however, still well suppressed compared with the
size of the gaugino mass, and so the usual gaugino mediation
spectrum results even for this two lattice site example.

One concern is that the link field might communicate SUSY
breaking to the MSSM matter scalars, since it is charged
under all gauge groups in this two site example.  We have already 
shown in Eq.~(\ref{Q1-mass-eq}) that the first link field acquires 
a SUSY breaking mass of order the gaugino (mass)$^2$.  However, there 
are no superpotential couplings between the link field and the matter 
scalars, so at most this field gives a (flavor diagonal) two-loop 
suppressed contribution to the MSSM matter scalars through loops
of the gauge and gaugino field.  This is suppressed by one more loops
than the contribution found above, and so can be neglected.

Some fine-tuning of $a^{-1} \ll \Lambda$ is needed in this case, 
however, to suppress the the two-loop gauge-mediated contribution 
(\ref{two-loop-simple-eq}).  In addition, operators like (\ref{link-op-eq})
could also lead to additional scalar mass contributions
\begin{eqnarray}
&&
\int  \, d^4 \theta \, \frac{S^\dag S}{\Lambda^{3}} Q_1 L_i^\dag L_j 
= \frac{v}{\Lambda} \left|\frac{F_S}{\Lambda}\right|^2
\phi_{L_i}^* \phi_{L_j} \; .
\end{eqnarray}
but they are suppressed by the same factor $v/\Lambda \sim 1/(\Lambda a)$
and so can be similarly suppressed.  In addition, we expect additional 
loop suppression of the coefficient of this operator when
the mediation of supersymmetry breaking occurs through messengers.
The hierarchy of scales 
that remains is $a^{-1} \sim v \ll \Lambda < 10^{-2} M_{\rm Pl}$
so that the contribution from the Planck suppressed operators 
are small.  This means new physics is appearing at scales
below the usual gauge coupling unification scale.

%%%%%%%%%%%%
\subsection{Gauge coupling unification}

Finally, we discuss the issue of unification of the gauge couplings
in this scenario.  There are two limiting cases that we will
consider, namely two gauge
groups, and alternatively, large $N$.  

For a larger number of lattice sites $N$ one does not need to separate the
scale of SUSY breaking and the lattice spacing anymore. One can take
$\Lambda \sim v \sim 10^{16}$ GeV, since the gauge-mediated contributions 
to the scalar masses can now be suppressed by the $1/N^2$ factor in 
(\ref{two-loop-simple-eq}), assuming that the operators in (\ref{link-op-eq}) 
also come with loop suppression factors. For $N$ gauge groups, one will
have $N$ KK modes appearing, with the first massive mode roughly at the 
scale $v/N$. For large $N$, this scale is well below $v$, and thus one
expects a phase of rapid power-law running of the gauge couplings to start
at around this scale. If we assume, that in addition to the gauge bosons
the Higgs fields also live at every lattice site (that is the Higgs is in
the bulk in the continuum limit), then one can use the results of
\cite{DDG} to show that despite the power-law nature of the running the
gauge couplings still unify, but at a scale below $M_{GUT}$, 
between $M_{GUT}/N$ and $M_{GUT}$, which for moderately large 
$5 \leq N \leq 50$ is still a relatively high scale close to $M_{GUT}$.
Thus the unification of couplings can still be maintained for the case of
large number of gauge groups as well, 
but it will happen at a scale below $M_{GUT}$, and with
non-perturbative values for the couplings due to the period of 
power-law running. Once one gets above the scale
$v$, the running switches back to a logarithmic running, since at this 
scale the gauge groups are not broken to the diagonal subgroup anymore.

For just two gauge groups
we have seen that $v \ll \Lambda \ll M_{\rm Pl}$, and so
the gauge couplings are not unified at the energy scale
where the lattice opens up.  To see what happens in this case,
we take the unbroken gauge group on each lattice site to be just 
SU(3) $\times$ SU(2) $\times$ U(1) and assume the three gauge 
couplings of {\it each} lattice site are the same for a given 
gauge group.  Above the scale $v$, 
the full 4D theory is the SM $\times$ SM, in which the MSSM matter 
is charged under one of them, while the link fields are charged under both.
We take three link fields, one transforming as a bifundamental 
under SU(3) $\times$ SU(3), one transforming as a bifundamental under
SU(2) $\times$ SU(2) and one field charged under U(1) $\times$ U(1).
These fields acquire scalar component vevs that break the SM $\times$ SM 
group structure down to just the SM.

The ordinary MSSM gauge couplings $\alpha_a$ for
$a=(U(1)_{\sqrt{3/5}Y}, SU(2), SU(3))$ are related to the gauge 
couplings $\overline{\alpha}_a$ of the endpoint lattice group through
$\alpha_a = \overline{\alpha}_a/2$, evaluated near the scale $v$.  
The gauge couplings therefore appear to undergo a discontinuous jump
at this scale, if we follow the gauge couplings of the diagonal subgroup 
up to $v$ and then the gauge couplings of the MSSM lattice site 
for energies beyond $v$.  We can evolve the individual gauge couplings 
$g_a$ above the scale $v$ by the usual renormalization group procedure.  
Of course we must also include the link fields (and the anomaly 
cancellation fields) in the beta functions.  At one-loop the running
of the gauge couplings from $\overline{\Lambda}_h$ down
to $v$ is given by the usual formula
\begin{equation}
\frac{1}{\overline{\alpha}_a(\overline{\Lambda}_h)} = 
\frac{1}{\overline{\alpha}_a(v)} 
- \frac{\overline{b}_a}{4 \pi} 
  \ln \frac{\overline{\Lambda}_h}{v}
\end{equation}
written entirely in terms of the parameters of the lattice
site gauge groups (barred quantities), where 
$\overline{\alpha}_a(\overline{\Lambda}_h,v)$ are the gauge
couplings for the two scales $\overline{\Lambda}_h,v$.
Let us now decompose this expression in terms of the usual
MSSM gauge couplings $\alpha_a$ and beta function coefficients $b_a$.
The diagonal subgroup will feel one KK mode of gauge and gaugino
fields, but since these fields have masses of order $v/2$, we can 
approximately treat this scenario as going from the 4D MSSM
lattice site directly to the diagonal subgroup with just the
4D MSSM matter content.  That is, $\overline{\alpha}_a(v) 
\rightarrow N \alpha_a(v)$.  The beta function coefficient 
above the scale $v$ is
\begin{equation}
\overline{b}_a \; = \; b_a + n_a 
               \; = \; \left( \frac{66}{5}, 2, -6 \right) 
                       + \left( 6/5, 4, 6 \right) \; .
\end{equation}
With the ratio $n_3/n_2 = 3/2$, we can choose the hypercharge of 
the link field connecting U(1) $\times$ U(1) such that the above 
shift in the beta function coefficients satisfies the 
conditions for one-loop unification given by Ref.~\cite{DDG}
\begin{equation}
\frac{B_{12}}{B_{23}} = \frac{B_{13}}{B_{23}} = 1 \qquad
B_{ij} = \frac{n_i - n_j}{b_i - b_j} \; .
\end{equation}
Hence, unification of gauge couplings can be maintained for the
$N=2$ case, although at a slightly lower scale than in the usual 4D MSSM.
However, in order to have a fully unified theory one would also have to 
embed these link fields into the GUT group.

%%%%%%%%%%%%%%%%%%%%%%%%%%%%%%%%%%%%%%%%%%%%%%%%%%%%%%
%%%%%%%%%%%%%%%%%%%%%%%%%%%%%%%%%%%%%%%%%%%%%%%%%%%%%%
\section{Conclusions}
%%%%%%%%%%%%%%%%%%%%%%%%%%%%%%%%%%%%%%%%%%%%%%%%%%%%%%
%%%%%%%%%%%%%%%%%%%%%%%%%%%%%%%%%%%%%%%%%%%%%%%%%%%%%%

We have presented 4D constructions for supersymmetric models
with extra dimensions. We have found that in the simplest model
(5D ${\cal N}=1$ SUSY YM) the necessary enhancement of 4D ${\cal N}=1$
supersymmetry automatically takes place without any fine-tuning, and
thus 5D Lorentz invariance is also recovered. For a theory with more
complicated matter content this result no longer holds, and a 
fine-tuning in the interaction terms is necessary. We have used these
models to translate the 
{\em a priori} five dimensional
mechanism of gaugino mediation of supersymmetry breaking into a simple 4D
model. In these 4D versions of gaugino mediation supersymmetry
breaking is transmitted to the MSSM because the physical gaugino
is a mixture of gauge eigenstate gauginos, one of which couples
to the supersymmetry breaking sector, while another to the SM matter
fields.  We find that a lattice as coarse as two gauge groups
is sufficient to ensure the appearance of the soft breaking mass spectrum
characteristic of gaugino mediation so long as the inverse lattice spacing
is much smaller than the SUSY mediation scale.  With more
than about five gauge groups one also obtains the gaugino-mediated
spectrum even if the SUSY mediation scale is of order the inverse 
lattice spacing.  

%%%%%%%%%%%%%%%%%%%%%%%%%%%%%%%%%%%%%%%%%%%%%%%%%%%%%%
%%%%%%%%%%%%%%%%%%%%%%%%%%%%%%%%%%%%%%%%%%%%%%%%%%%%%%
\section*{Acknowledgements}
%%%%%%%%%%%%%%%%%%%%%%%%%%%%%%%%%%%%%%%%%%%%%%%%%%%%%%
%%%%%%%%%%%%%%%%%%%%%%%%%%%%%%%%%%%%%%%%%%%%%%%%%%%%%%
We are extremely grateful to Tanmoy Bhattacharya for his continued 
help with lattice gauge theories and for collaboration at early stages 
of this work. We also thank Nima Arkani-Hamed, Sasha Gorsky, 
Weonjong Lee, Martin Schmaltz, Yuri Shirman and Raman Sundrum
for useful discussions, and Nima Arkani-Hamed, Andy Cohen,
Stefan Pokorski and Yuri Shirman for valuable
comments on the manuscript. C.G. and G.D.K. thank the members of the T-8
group at Los Alamos for their hospitality while this work was
initiated. 

C.C. is an Oppenheimer fellow at the Los Alamos National Laboratory.
This research of C.C. and J.E. is supported by
the U.S. Department of Energy under contract W-7405-ENG-36. C.G. is 
supported in part by the US Department of Energy under contract
DE-AC03-76SF00098 and in part by the National Science Foundation under
grant PHY-95-14797. 
G.D.K. is supported in part by the U.S. Department of Energy 
under contract DE-FG02-95-ER40896.

%%%%%%%%%%%%%%%%%%%%%%%%%%%%%%%%%%%%%%%%%%%%%%%%%%%%%%
%%%%%%%%%%%%%%%%%%%%%%%%%%%%%%%%%%%%%%%%%%%%%%%%%%%%%%
%\end{document}

\end{document}